\documentclass[10pt,journal,compsoc]{IEEEtran}
\ifCLASSOPTIONcompsoc
  \usepackage[nocompress]{cite}
\else
  \usepackage{cite}
\fi
\ifCLASSINFOpdf
\else
\fi
\usepackage{graphicx}
\usepackage{amsmath}
\usepackage{multirow}
\usepackage{amsfonts,amsmath}
\usepackage{amssymb, nccmath}
\usepackage{amssymb, nccmath}
\usepackage{dsfont}
\usepackage{booktabs}
\usepackage[utf8]{inputenc}
\usepackage{dsfont}
\usepackage{booktabs}
\usepackage{graphicx}
\usepackage{grffile}
\usepackage{subfigure} 
\usepackage{grffile}
\usepackage{array} 
\usepackage{mathtools}
\usepackage{fontenc}
\usepackage{tikz}
\usepackage{pgfplots}
\usepackage{tikz}
\usepackage{verbatim}
\usetikzlibrary{patterns}
\usepackage[linesnumbered, ruled]{algorithm2e}
\makeatletter
\newcommand{\removelatexerror}{\let\@latex@error\@gobble}
\def\ps@IEEEtitlepagestyle{%
	\def\@oddfoot{\mycopyrightnotice}%
	\def\@oddhead{\hbox{}\@IEEEheaderstyle\leftmark\hfil\thepage}\relax
	\def\@evenhead{\@IEEEheaderstyle\thepage\hfil\leftmark\hbox{}}\relax
	\def\@evenfoot{}%
}

\def\mycopyrightnotice{%
	\begin{minipage}{\textwidth}
		\centering \scriptsize
		This article has been accepted in IEEE Transactions on Parallel and Distributed Systems Journal © 2021 IEEE. Personal use of this material is permitted. Permission from
		IEEE must be obtained for all other uses, in any current or future media, including reprinting/republishing this material for advertising or promotional purposes, creating new collective works, for resale or redistribution to servers or lists, or reuse of any copyrighted component of this work in other works. This work is freely available for survey and citation.
		
	\end{minipage}
}
\makeatother
\usepackage{subfig}

\renewcommand\thesection{\arabic{section}}

\renewcommand{\thesubsubsection}{\Alph{subsubsection}}
\begin{document}
%
\title{ A Quantum Approach Towards the Adaptive Prediction of Cloud Workloads}
%
%
%
%
\author{Ashutosh~Kumar~Singh,~\IEEEmembership{Senior member IEEE},~Deepika~Saxena,~\IEEEmembership{}~Jitendra~Kumar,~and~Vrinda~Gupta
\IEEEcompsocitemizethanks{\IEEEcompsocthanksitem  D. Saxena and A. K. Singh are co-first authors. They are with the Department of Computer Applications, NIT Kurukshetra\\E-mail: 13deepikasaxena@gmail.com,  ashutosh@nitkkr.ac.in\\ J.Kumar is with the Department of Computer Applications, NIT Tiruchirappalli, Tamilnadu, India. E-mail: jitendra@nitt.edu\\ V.Gupta is with the Department of Electronics \& Communication Engineering, NIT, Kurukshetra, India. E-mail: vrindag16@nitkkr.ac.in 
}}
\markboth{IEEE TRANSACTIONS ON PARALLEL AND DISTRIBUTED SYSTEMS}%
{Shell \MakeLowercase{\textit{et al.}}: Bare Demo of IEEEtran.cls for Computer Society Journals}
\IEEEtitleabstractindextext{%
\begin{abstract}
This work presents a novel Evolutionary Quantum Neural Network (EQNN) based workload prediction model for Cloud datacenter. It exploits the computational efficiency of quantum computing by encoding workload information into qubits and propagating this information through the network to estimate the workload or resource demands with enhanced accuracy proactively. The rotation and reverse rotation effects of the Controlled-NOT (C-NOT) gate serve activation function at the hidden and output layers to adjust the qubit weights. In addition, a Self Balanced Adaptive Differential Evolution (SB-ADE) algorithm is developed to optimize qubit network weights. The accuracy of the EQNN prediction model is extensively evaluated and compared with seven state-of-the-art methods using eight real world benchmark datasets of three different categories. Experimental results reveal that the use of the quantum approach to evolutionary neural network substantially improves the prediction accuracy up to 91.6\% over the existing approaches.   

\end{abstract}

}

\maketitle

\IEEEdisplaynontitleabstractindextext

\renewcommand\thesubsection{\Alph{subsection}}

%

\IEEEraisesectionheading{\section{Introduction}\label{sec:introduction}}

\ifCLASSOPTIONcompsoc
\IEEEPARstart{C}{loud} computing has emerged as an indispensable paradigm for computation and storage over the internet which is widely used in business, marketing, online transactions, research, academia, etc. \cite{buyya2018manifesto}. It provides ubiquitous computing services benefit to all its users and freedom to pay-as-per-use model \cite{saxena2021op}, \cite{saxena2020energy}. Cloud service providers (CSP) utilize virtualization \cite{song2013adaptive}, \cite{saxena2020security} of physical resources at datacenters to maximize their revenue. The comprehensive working of datacenters depend on fine-grained provisioning of resources like storage, processor and network etc. \cite{kumar2018workload}, \cite{singh2021cryptography}. The workload demands show high variation over the time causing over/under-load and SLA violation problems where the static allocation of resources is unapplicable \cite{bi2019temporal}. Evidently, there is a critical requirement of an efficient forecasting system that can predict upcoming resource utilization and workload demands with utmost precision to allow flawless functioning of cloud datacenters. The predicted workload information helps to analyze the right amount of resources to scale over a particular time interval, improve resource utilization, reliability (through prior prediction of system failures), minimize power consumption and make resource management decisions \cite{saxena2021op}.   However, the accurate workload prediction is highly challenging due to the fluctuations in resource demands which has grabbed the attention of numerous researchers across the globe.
\par 
The classical workload forecasting methods based on time-series analysis, regression, and heuristic theories, artificial neural networks perform better for workloads with regular patterns. Meanwhile, the conventional neural network does not fully exploit the correlation between extracted patterns for accurate prediction \cite{chen2019towards}. It has been observed that quantum bits can draw more precise correlations from complex and dynamic input samples as compared to real numbers, which are used to set-up conventional neural network \cite{ezhov2001role}. Hence, the learning capability of a conventional neural network can be enhanced by applying computational properties of quantum computing \cite{cutting1999would}. Quantum superposition functionality can be derived in the form of quantum states or qubit phase values as an operational unit by the quantum neural network (QNN) model \cite{biamonte2017quantum}. Qubit phase values are generated from rotational effect, which have richer capability to learn more complex relationships during training process \cite{ezhov2001role}. This gives the motivation for proposing a quantum approach based neural network prediction model for dynamic workload predictions. The most common method applied for training of conventional  neural network is back propagation that employs single solution and tries to adjust it during learning process. However, it suffers from getting stuck into local minima and pre-mature convergence. On the other hand, evolutionary algorithms explore number of solutions to search an optimal solution and provide derivative-free optimization of neural network. As compared to GA and PSO, differential evolution (DE) has greater optimization capabilities in terms of convergence speed, computational complexity, accuracy and stability \cite{hegerty2009comparative}. Therefore, a Self Balanced and Adaptive Differential Evolution (SB-ADE) algorithm is developed with self balancing capabilities of exploration and exploitation for optimization of proposed quantum approach based neural network prediction model.
{The key contributions are:}
\begin{itemize}
	\item A novel \textbf{E}volutionary \textbf{Q}uantum \textbf{N}eural \textbf{N}etwork (\textbf{EQNN}) based workload prediction model is proposed to proactively estimate the dynamic resource and job arrival demands with improved accuracy at cloud datacenters.
	\item SB-ADE learning algorithm is developed, which explores the search space globally and exploits the population to move closer towards a better solution to optimize the EQNN model.
	\item {In-depth experimental analysis of the prediction accuracy of EQNN model is carried out over eight real world benchmark datasets belonging to three different categories. The comparison of EQNN prediction model with seven state-of-the-art methods confirms its superior accuracy.}
	\end{itemize}
\par The rest of the paper is organized as: Section 2 discusses the related work. The proposed approach is described in Section 3, which is divided into three subsections including, EQNN Prediction Model, its information processing, and training by SB-ADE algorithm. Section 4 entails performance evaluation followed by conclusive remarks and future scope of the proposed work in Section 5. 
\section{Related Work}
The related work covers neural network based workload prediction and quantum inspired information processing.
\subsection{Workload prediction}
A future workload prediction technique based on Back-propagation i.e., Random Variable Learning Rate Back-propagation Neural Network (RVLRBPNN) was developed in \cite{lu2016rvlbpnn} to forecast the number of requests expected per prediction-interval. Later, Kumar et al. have proposed a self-adaptive differential evolution (SaDE) based artificial neural network model \cite{kumar2018workload} for workload prediction which is capable of producing  more accurate results as compared to RVLRBPNN gradient decent based backpropagation trained neural network model of \cite{lu2016rvlbpnn}. {Also, neural network based workload predictors trained with different evolutionary algorithms were presented in \cite{saxena2020auto}, \cite{kumar2020biphase}, have outperformed the accuracy achieved by Back-propagation training based neural network. This was due to multi-directional searching with better exploration and learning capability of evolutionary algorithms over Back-propagation}. The long short-term memory model in a recurrent neural network (LSTM-RNN) for fine-grained host load prediction was presented in \cite{kumar2018long}. Though the LSTM-RNN model learns long-term dependencies and produces high accuracy for the prediction of server loads, it suffers from long computation time during training due to the usage of Back-propagation algorithm between recurrent layers. To allow high capability of learning and better accuracy in less time, multi-layered neural networks with multi-valued neurons (MLMVN) prediction model was proposed in {\cite{qazi2018cloud}}. This approach used complex-valued neural network with a derivative-free feed-forward learning algorithm and produced better accuracy than LSTM-RNN approach. Recently, a neural network based multi-resource predictor is proposed in \cite{saxena2020proactive} which is trained using an adaptive evolutionary algorithm to predict multiple resources simultaneously. Table \ref{summaryNN} compares major features of EQNN and state-of-the-art workload prediction approaches. 
\begin{table}[htbp]
	
	\caption{{Major features: EQNN vs state-of-the-arts}}
	\label{summaryNN}
	\scriptsize
	\centering
	\resizebox{0.9\textwidth}{!}{\begin{minipage}{\textwidth}
			\begin{tabular}{p{1.9cm}  p{1.9cm}   p{1.7cm} p{1.0cm} c c}
			{	\textbf{Models}}&{\textbf{Training Algorithm}}&{\textbf{Datasets}}&{\textbf{Error metrics}}&{\textbf{AEF$^a$}}& {\textbf{SA$^b$}} \\ \hline
			
		{	SaDE+NN \cite{kumar2018workload}}&{ Self adaptive DE} &{NASA and Saskatchewan}  &{RMSE}&{$\times$}&{$\times$} \\ \hline
		{	{RVLRBPNN \cite{lu2016rvlbpnn} }}& {Backpropagation }&{NASA } &{RMSE}&{$\times$}&{$\times$} \\ \hline
		{	AaDE+NN  \cite{saxena2020auto}}& {DE with three-phase adaptation}&{NASA and Saskatchewan}  &{RMSE}&{$\times$}&{$\checkmark$} \\ \hline
			{BaDE+NN  \cite{kumar2020biphase}}&{Two-phase adaptive DE} &{NASA and Saskatchewan } &{RMSE}&{$\times$}&{$\times$} \\ \hline
		
			{LSTM-RNN \cite{kumar2018long}} &{Backpropagation} &{Google Cluster and HPC/Grid System}  &{MSE, MMSE}&{$\times$}&{$\times$} \\ \hline
			 {MLMVN  \cite{qazi2018cloud}}&{Backpropagation} &{Sahara, Google Cluster}  &{MSE}&{$\times$}&{$\times$} \\ \hline
			{OM-FNN \cite{saxena2020proactive}} & {Variant of DE algorithm} &{Google Cluster}  &{RMSE}&{$\times$}&{$\times$ }\\ \hline
		{\textbf{EQNN}} &{SB-ADE} &{ Cluster:3, HPC:3, Web:2} &{Normalised RMSE, MAE}&{$\checkmark$}&{$\checkmark$} \\ \hline
			\end{tabular}
	\end{minipage}}
\footnotesize{\tiny{{$^a$ AEF:Absolute Error Frequency, $^b$ SA:Statical Analysis}}}
\end{table}

\subsection{Quantum based Information processing}
Many researchers have established that quantum computing can be assembled with machine learning through the concept of quantum parallelism \cite{havenstein2018comparisons}, \cite{imre2007quantum}. For instance, Quantum based fog scheduler is presented in \cite{bhatia2019quantum} to determine optimal fog node for data analysis in real time. Quantum inspired information processing is based on qubits which allow faster learning and pattern recognition ability for complex relations between specific patterns of input data samples \cite{zhou2007quantum}. {A quantum back propagation (QBP) neural network model was proposed in \cite{matsui2000neural} for solving the 4-bit parity check problem  that performed excellently in information processing efficiency than a classical neural network. Later, in \cite{kouda2005qubit}, the learning capability of Qubit Neuron Network on 4-bit and 6-bit parity check problems have been investigated and result entailed that Qubit NN was far better than classical NN and complex NN that was due to the implication of superposition of quantum states at neuron node and the probability interpretation. Kouda et al. proposed multi-layer feed-forward QNN model \cite{kouda2004multilayered} and analyzed it for high‐level information processing, which revealed that QNN produced much better results than conventional Neural Network. 
The Qubit Multi-layer Perceptron (QuMLP) with a Quantum-Inspired Evolutionary Algorithm (QIEA) can search for the best specific time lags which is able to characterize the time series phenomenon, and can evolve the complete QuMLP architecture and parameters to control the random walk dilemma problem and has been analyzed for financial time series prediction in \cite{araujo2010quantum}. Cui et al. proposed a Complex Quantum Neuron (CQN) model for the time series prediction and the results improved due to realization of a deep quantum entanglement \cite{cui2015complex}. Ganjefar et al. \cite{ganjefar2017training} proposed a quantum states based neural network optimized with hybrid genetic algorithm and gradient descent (HGAGD) studied the feasibility of designing an indirect adaptive controller for damping of low frequency oscillations in power systems and achieved satisfactory control performance.} Table \ref{summaryQNN} summarises the operation and performance of QNN across the existing application domain.
\par Unlike existing works which have utilized traditional machine learning models and optimization algorithms for cloud workload prediction, the proposed work exploits the computational ability of quantum principles with machine learning capabilities of evolutionary neural network. Since the quantum systems can produce counter intuitive behavioural patterns, that are believed not to be efficiently produced by classical systems, it is reasonable to postulate that quantum computation combined with neural network optimization model outperforms the classical machine learning approaches in terms of its adaptability to forecast wider range of workloads with enhanced prediction accuracy.

\begin{table}[htbp]

	\caption{{Usage of QNN across the application domain}}
		\label{summaryQNN}
	\scriptsize
	\centering
	\resizebox{0.9\textwidth}{!}{\begin{minipage}{\textwidth}
			\begin{tabular}{p{1.5cm}  p{1.5cm}   p{2.0cm} p{2.5cm} }
				{\textbf{Models}}&{\textbf{Underline Algorithm}}&{\textbf{Application}}&{\textbf{Performance metrics} }\\ \hline
				
			{	Quantum computing inspired NN \cite{bhatia2019quantum}}& {Gradient Descent} & {Prediction based Fog scheduler}  &{task completion time and average energy consumption}
				\\ \hline
				{Qubit NN \cite{kouda2005qubit}}& {Backpropagation}&{4-bit parity check problem}&{learning efficiency in terms of iterations}\\ \hline
				{QuMLP\cite{araujo2010quantum}}&{QIEA}&{stock prediction}&{MSE, MAPE}\\ \hline
				{CQN \cite{cui2015complex}}&{deep quantum entanglement}&{time series prediction}&{MSE, MAPE}\\ \hline
				{Quantum inspired NN\cite{ganjefar2017training}}& {HGAGD}&{designing adaptive controller in power system}&{Integral Absolute Error, Integral Square Error,
				Integral Time Square Error}\\ \hline
			\end{tabular}
	\end{minipage}}
\end{table}
\section{Proposed approach}
{The proposed workload prediction approach consists of three phases: \textit{Learning}, \textit{Testing}, and \textit{Prediction} as shown in Fig. \ref{fig:loadpredictionworkflow}. During learning, the historical data is pre-processed by applying attribute extraction and normalization, followed by its division into training, validation, and testing data. The EQNN extracts suitable patterns from training data and finds out correlations among them to generate EQNN prediction trained model. The validation data is used to tune or optimize the model by evaluating its accuracy using an error evaluation function. If the desired accuracy is not achieved, the prediction model is retrained; otherwise, the model enters into the testing phase.} During testing, the accuracy of the trained model is evaluated with unseen/test data to generate final prediction model. Finally, in load prediction phase, the EQNN prediction model is deployed for a proactive estimation of cloud workload and help in proficient load management. A detailed description of proposed approach is given in the subsequent sections.  

\begin{figure}[!htbp]
	\centering
	\includegraphics[width=0.99\linewidth]{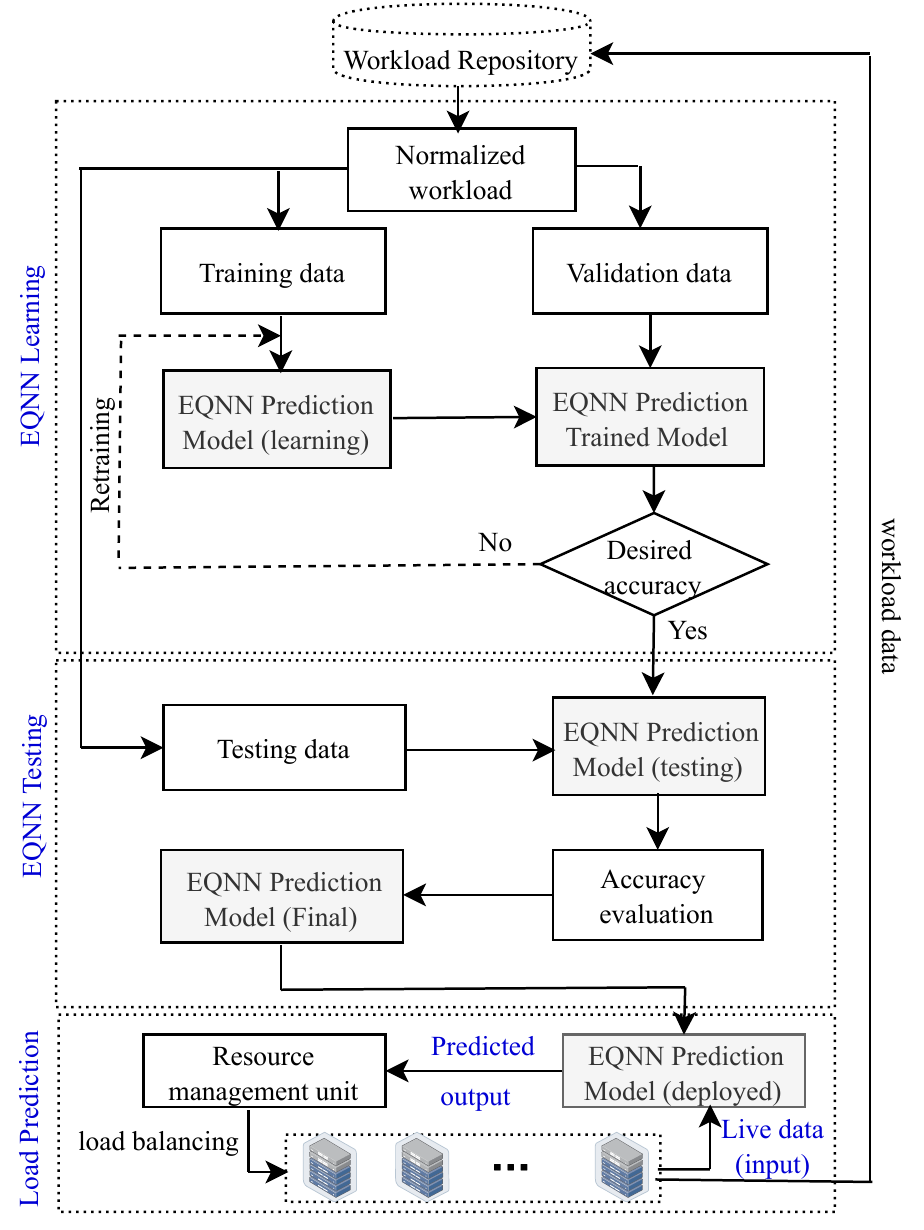}
	\caption{Load Prediction Workflow}
	\label{fig:loadpredictionworkflow}
\end{figure}

\subsection{ EQNN Prediction Model}
 EQNN is an intelligent prediction model that utilizes the machine learning ability of evolutionary neural network and computational efficiency of quantum principles to maximize prediction accuracy. The fundamental unit of QNN is qubit, which comprises of two-state quantum-mechanical system. A single qubit can represent a one, a zero, or crucially, any quantum superposition of these. Mathematically, a qubit can be realized as $|\Psi\textgreater = \alpha|0\textgreater +  \beta|1\textgreater,    $ 
where $\alpha$ and $\beta$ are complex numbers specifying the probability amplitudes of states $|0\textgreater$ and $|1\textgreater$ respectively. 

\par The proposed EQNN prediction model shown in Fig. \ref{fig:updatedQNN} consists of three layers, including input, hidden and output layers having $n$, $p$ and $q$ qubit nodes respectively which represents $n$-$p$-$q$ qubit network architecture. The connection weights between qubit neurons of different layers are also taken in the form of qubits that are adjusted during training of EQNN. 
\begin{figure*}[htbp]
	\centering
	\includegraphics[width=0.8\linewidth]{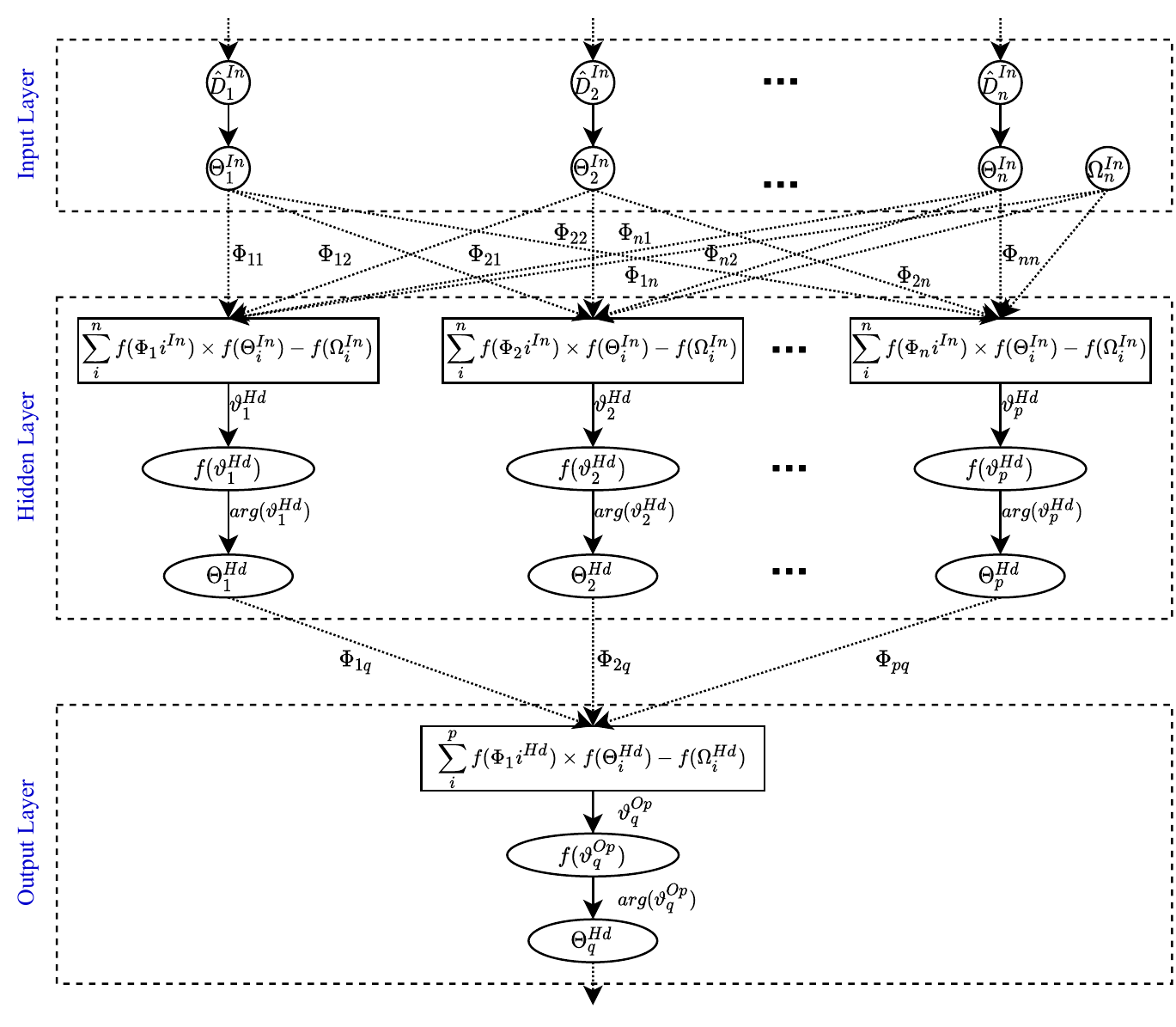}
	\caption{Evolutionary Quantum Neural Network (EQNN) Model for Workload Prediction}
	\label{fig:updatedQNN}
\end{figure*}
The qubit neuron state transitions are built upon the operations derived from quantum logic gates, including quantum rotation and C-NOT gates. The rotational effect of the quantum rotation gate is applied to generate qubits, and the rotation and reverse rotation functionality of the C-NOT gate is applied as an activation function at the hidden and output layers \cite{dos2019quantum}. The quantum rotation or phase-shift gate transforms the phase of quantum state or qubit which is represented in Eq. \ref{eq3}:
\begin{equation}\label{eq3}
R(\phi)= 
\left[ {\begin{array}{cc}
	cos\phi & -sin\phi\\
	sin\phi & cos\phi\\
	\end{array} } \right]
\end{equation}
Consider initial quantum state $|\Psi\textgreater=|cos\phi_0, sin\phi_0|$, where $|\Psi\textgreater$ can be transformed by applying $R(\phi)$ that {shifts the phase} of $|\Psi\textgreater$ as depicted in Eq. \ref{qubit}.
\begin{equation}
\label{qubit}
R(\phi)|\Psi\textgreater= 
\left[ {\begin{array}{cc}
	cos\phi & -sin\phi\\
	sin\phi & cos\phi\\
	\end{array} } \right]
\left[ {\begin{array}{c}
	cos\phi_0 \\
	sin\phi_0\\
	\end{array} } \right]
=\left[ {\begin{array}{c}
	cos(\phi +\phi_0) \\
	sin(\phi +\phi_0)\\
	\end{array} } \right]
\end{equation}\textit{Controlled-NOT gate} 
The C-NOT gate \cite{kouda2005qubit} performs reverse rotation and non-rotation of quantum states  with respect to controlled input parameter $\eta$ and can be defined as shown in Eq. \ref{eq4}, where $\eta=1$ and $\eta=0$ stands for reverse rotation and non-rotation, respectively.
\begin{equation}\label{eq4}
f(\frac{\pi}{2}\times \eta  - \phi)=
\begin{cases}
sin(\phi) + icos(\phi), & {If(\eta=1)} \\
cos(\phi) - isin(\phi), & {If(\eta=0)} 
\end{cases}
\end{equation}

\subsubsection{Data pre-processing and Qubit generation}

 The training input values (historical values) are extracted and aggregated into a specific time-interval (for example, 1 min, 5 min, 10 min and so on). The aggregation step is followed by the normalization of input data $D$ in the range [0, 1] using {$\hat{D}= \frac{ D_i- D_{min}}{D_{max}-D_{min}}$, where $D_{min}$ and $D_{max}$ are the minimum and maximum values of the input data set, respectively}. The normalized vector is denoted as $\hat{D}$, which is a set of all normalized input data values \{$\hat{D_1}$, $\hat{D_2}$, ..., $\hat{D_n}$\}. The normalized data is fed into the input layer of EQNN to transform the real-valued inputs into the quantum state phase values in the range $(-\pi/2, +\pi/2)$ by applying the effect of qubit rotation using ${y^{In}}_i=f({\Theta_i}^{In})$ and $\Theta_i=\frac{\pi}{2}\times d_i $, where $d_i$ is the $i^{th}$ input data point, $\Theta_i$ is $i^{th}$ quantum input to the network.
   
These normalized values (in single dimension) are arranged as  multi-dimensional qubit-input matrix and qubit-output matrix denoted as $\Theta_{input}$ and $\Theta_{output}$ respectively as stated in Eq. \ref{eqn:7}:
\begin{equation}
\resizebox{0.5\textwidth}{!}{$  
	\label{eqn:7}
	\Theta_{input}= 
	\left[ {\begin{array}{cccc}
		\Theta_1 & \Theta_2 & .... & \Theta_n\\
		\Theta_2 & \Theta_3 & .... & \Theta_{n+1} \\
		.    &    .     & .... &    .     \\
		.    &    .     & .... &    .     \\
		\Theta_m & \Theta_{m+1} & .... & \Theta_{n+m-1} \\    
		\end{array} } \right]
	\Theta_{output}= 
	\left[ {\begin{array}{c}
		\Theta_{n+1} \\
		\Theta_{n+2}  \\
		.      \\
		.        \\
		\Theta_{n+m}    
		\end{array} } \right]$}
\end{equation}

The QNN model extracts intuitive patterns from actual workload ($\Theta_{input}$) and analyzes $n$ previous workload values to predict the workload ($\Theta_{output}$) about to arrive at next $(n+1)^{th}$ instance of time at the datacenter.
\subsection{QNN Information Processing}
The operational QNN model developed in the current approach is shown in Fig. \ref{fig:updatedQNN}. The Input layer neurons receive data as real valued data input \{$\hat{D_1}$, $\hat{D_2}$, ..., $\hat{D_n}$\} and transform them into a set of qubit state vector such as $ \{\Theta^{In}_1, \Theta^{In}_2,...,\Theta^{In}_n\}$ by applying the steps mentioned in Qubit generation phase. In addition, a threshold qubit value ($\Omega^{In}$) is generated in the range [$(-\pi/2, +\pi/2)$] as a bias. The input layer is connected to the hidden layer through qubit neural weights represented as $\Phi^{In}_{11}$, $\Phi^{In}_{12}$,..., $\Phi^{In}_{np}$. 

\par At the Hidden layer, qubit vector such as $ \{\Theta^{Hd}_1, \Theta^{Hd}_2,...,\Theta^{Hd}_p\}$, is generated by computing $ y_j^{Hd}=f(\Theta_j^{Hd})$ using Eqs. (\ref{hidden2} and \ref{hidden3}), where $y_j^{Hd}$ is the $j^{th}$ output of hidden layer, and  $f(\Theta_j^{Hd})$ states that an activation function is applied to $\Theta_j^{Hd}$.

\begin{gather}
 \Theta_j^{Hd} = \frac{\pi}{2}\times g(\delta^{Hd}) - arg(\vartheta_j^{Hd}) \label{hidden2}\\   
\vartheta_j^{Hd}=\sum_{i=1,j=1}^{n,p}f(\Phi_{ij}^{In})\times f( \Theta_i^{In})-f(\Omega^{In})  \label{hidden3}   
\end{gather}
where $\Theta_j^{Hd}$ is $j^{th}$ (adjusted) qubit vector, obtained by applying reverse rotation or non-rotation effect of C-NOT gate, on amplitude (or magnitude) of qubit vector represented as $arg(\vartheta_j^{Hd})$, generated by applying activation function to ${\vartheta_j^{Hd}}$ obtained from Eq. \ref{hidden3}. The expression $g(\delta^{Hd})$ is sigmoid function that produces value in the range [0, 1]. The Output layer generates prediction output as $y^{Op}$, which is derived by computing $y^{Op}=f(\Theta^{Op})$. Similar to the hidden layer, $\Theta^{Op}$ is a list of (adjusted) qubits, obtained by applying C-NOT gate effect on the amplitude of intermediate qubit vector i.e., $arg(\vartheta^{Op})$ as stated in Eq. \ref{output2}. The intermediate qubit vector $\vartheta^{Op}$ can be obtained by applying Eq. \ref{output3}.
\begin{gather}
\Theta^{Op} = \frac{\pi}{2}\times g(\delta^{Op} ) - arg(\vartheta^{Op}) \label{output2}\\
\vartheta^{Op}=\sum_{i=1}^{p}f(\Phi_{i}^{Hd})\times f( \Theta_i^{Hd})\label{output3}
\end{gather}
The $sigmoid()$ function shown in Eq.\ref{sigmoidfunction} imparts non-linearity and maps input to output in the range [0, 1]. Thereby, predicted qubit values are smoothen in the range [0, 1] to make them comparable with the actual output values. Finally, the performance and accuracy of the proposed model are evaluated by applying the root mean square error score (acting as fitness function) given in Eq.\ref{rmse}. 
\begin{gather}
sigmoid(\varphi)=\frac{1}{1+exp^{(-\varphi)}}\label{sigmoidfunction}\\
RMSE = \sqrt{\frac{1}{m}\sum_{i=1}^{m}(z_{actual}-z_{prediction})^2}\label{rmse} 
\end{gather}

\subsection{Training by SB-ADE Algorithm}
The training of EQNN begins with a randomly generated set of $N$ qubit state vectors, where each vector represents a network of neural weight connections (computed using Eq. \ref{eqn:qq}), of size $(n+1)\times p + (p\times q) = p(n+q+1) \Leftarrow p(n+2)$, where $q=1$ and additional $1$ is added as bias at input layer.  
\begin{gather}
\label{eqn:qq}
\Psi=\alpha + i\beta \\
\Phi= phase(\Psi)\times \frac{\pi}{2} 
\end{gather} 
where $\alpha=sqrt(rd)$ and $\beta=sqrt(1-rd)$ and $rd=random(0,1)$. The term $rd$ is a randomly generated number in range [0, 1], $\alpha$ and $\beta$ are the probability amplitudes for realizing $|0\textgreater$ and $|1\textgreater$ respectively, $\Psi$ is a quantum state in the Hilbert vector space and $\Phi$ is the phase angle determined for the corresponding qubit.
 
The number of maximum generations (epochs), mutation and crossover rate vectors are initialized within a specific range. Each network is evaluated on training data by applying a fitness function (Eq. \ref{rmse}). The mutation strategy selection parameter ($msp$) is generated for each generation to select one of the four optional mutation schemes. The mutation and crossover operators are applied to generate offspring for next generation. The offspring vectors are evaluated by applying the fitness function which selects the optimal solutions for next generation. The mutation and crossover rates are updated after a fixed number of epochs ("learning period").  

\subsubsection{Mutation}
SB-ADE optimization algorithm is developed with four different mutation schemes that {balance} the exploration and exploitation effects with adaptive generation of mutation learning rate ($lp^M$). The adoption of mutation strategy plays an important role in improving the quality of solutions. Each mutation strategy has some characteristics associated with it that justifies its specific usage. For example, $DE/best/1$ (Eq. \ref{eqn:mutation1}) and $ DE/current-to-best/1 $ (Eq. \ref{eqn:mutation2}) mutation strategies tend to be greedy as they use the best individual to generate mutant vectors which brings comprehensive exploitation during evolutionary stages of the algorithm. The mutation strategies $DE/current-to-rand/1/$ (Eq. \ref{eqn:mutation3}) and $DE/rand/1$ (Eq. \ref{eqn:mutation4}) {help} to find new search direction randomly that prevents premature convergence by allowing exploration \cite{iorio2004solving} and raise population diversity that contributes to intensive exploration under control parameter $\kappa$. This will help in maintaining a good balance between exploration and exploitation properties of mutation variants.

\begin{equation} 
 \omega_i^j = \Phi_{best}^j + \mu_i \times (\Phi_{r1}^j -\Phi_{r2}^j)\label{eqn:mutation1}
\end{equation}

\begin{equation}
\omega_i^j = \Phi_i^j +  \mu_i \times (\Phi_{best}^j -\Phi_i^j) +  \mu_i \times (\Phi_{r1}^j -\Phi_{r2}^j)\label{eqn:mutation2}
\end{equation}

\begin{equation}
\omega_i^j = \Phi_i^j +  \kappa_i \times (\Phi_{r1}^j -\Phi_i^j) +  \mu_i \times (\Phi_{r2}^j -\Phi_{r3}^j)\label{eqn:mutation3}
\end{equation}
\begin{equation}
\omega_i^j = \Phi_{r3}^j + \mu_i \times (\Phi_{r1}^j -\Phi_{r2}^j) \label{eqn:mutation4}
\end{equation}
where $\omega_i^j$ and $\Phi_i^j$ depict $i^{th}$ mutant and current vectors respectively of $j^{th}$ iteration. The randomly generated numbers $r1$, $r2$ and $r3$ are mutually distinct numbers in the range [1, N]. The term $\Phi_{best}^j$ represents the best solution found so far, till $j^{th}$ generation. The parameters controlling mutation rate are $\mu_i$ and $\kappa_i$, for corresponding vector in $i^{th}$ iteration. Let $\Gamma_1$, $\Gamma_2$, $\Gamma_3$ and $\Gamma_4$ be the probabilities for opting the $DE/rand/1$, $ DE/best/1 $, $DE/current-to-best/1$ and $ DE/current-to-rand/1 $ mutation techniques, respectively. In reported experiments, initially $\Gamma_1$=$\Gamma_2$= $\Gamma_3$=$\Gamma_4$= 0.25, so that each mutation scheme gets an equal chance of selection. Roulette wheel selection \cite{zhang2012equal} scheme is utilized to choose among the mutation strategies that are based on the assigned probabilities. To implement the roulette wheel selection, a mutation selection probability ($msp$) vector of $N$ random numbers in the range [0, 1], is generated. For example, if $msp_i$ is less than or equal to  $ \Gamma_1 $, $DE/rand/1$ is applied. Similarly, if the $msp_i $ is greater than $ \Gamma_1 $ and smaller or equal to $\Gamma_1 + \Gamma_2 $, $DE/best/1$ is applied, and so on. The mutation strategy selection is represented as $\wp $ in Eq. \ref{mutation_probability}.
\begin{equation}\label{mutation_probability}
\resizebox{0.5\textwidth}{!}{$  
	\wp =
	\begin{cases}
	$DE/current-to-best/1$, & {If(0< msp_i \leq \Gamma_1 )} \\
	$ DE/best/1$ , & {If(\Gamma_1  < msp_i  \leq \Gamma_1 + \Gamma_2 )} \\
	$DE/rand/1$ , & {If(\Gamma_1 + \Gamma_2  < msp_i \leq \Gamma_1 + \Gamma_2 + \Gamma_3 )}  \\
	$  DE/current-to-rand/1 $ , & {\text{otherwise}} 
	\end{cases}$}
\end{equation}
\subsubsection{Crossover}

The crossover operation is applied to mutant vector $\omega_i^j$, and its current target vector $\Phi_i^j$ to produce new solutions called offspring $\chi_i^j$ i.e., $i^{th}$ solution of $j^{th}$ generation. A uniform crossover operation is applied to swap information at the gene level instead of segment level \cite{pavai2017survey}. Eq. \ref{crossover} shows a crossover operation where $\Re$ is a randomly generated number in the range [0, 1] for each gene in the chromosome. If the value of $\Re$ is smaller than the corresponding $CR_{vi}^j$, then the crossover is successfully applied to exchange $i^{th}$ gene of the mutant vector with the current target vector.
\begin{gather}\label{crossover}
\chi_i^j=\begin{cases}
\omega_i^j  & {If(\Re\in(0,1) \leq CR_{vi}^j )} \\
\Phi_i^j   & {\text{otherwise.}} 
\end{cases}
\end{gather}
In the proposed approach, the crossover rate is initialized randomly in the range [0, 1] with mean value $CR_\mu$ as 0.5, and standard deviation $CR_\sigma$ as 0.1. The crossover rate gets update after a fixed number of generations (known as crossover learning period) with new values of the mean and standard deviation of crossover rate which are recorded during evolutionary process.

During each generation or epoch, the number of candidates successfully reaching the next generation denoted as $\xi_1$, $\xi_2$, $\xi_3$, and $\xi_4$, for four different mutation strategies, are monitored. Similarly, $\Delta_1$, $\Delta_2$, $\Delta_3 $, and $\Delta_4$ record the number of candidates who failed to reach the next generation.
The probabilities for opting $DE/rand/1$, $ DE/best/1 $, $DE/current-to-best/1$, and $ DE/current-to-rand/1 $ mutation strategies are computed as  $ \Gamma_1 $,  $ \Gamma_2 $,  $ \Gamma_3 $, and  $ \Gamma_4 $ by applying Eq. \ref{eqn:probability}.

\begin{equation}\label{eqn:probability}
\begin{gathered}
Z= 2(\xi_2\xi_3\xi_4 + \xi_1\xi_3\xi_4 + \xi_1\xi_3\xi_2 + \xi_2\xi_3\xi_4) + \Delta_1(\xi_2 + \xi_3+\xi_4) + \\ \Delta_2(\xi_1 + \xi_3+\xi_4) + \Delta_3(\xi_1 + \xi_2+\xi_4) + \Delta_4(\xi_1 + \xi_3+\xi_2) \\
(\Gamma_1)= \frac{\xi_1(\xi_2 + \Delta_2 + \xi_3 + \Delta_3 +\xi_4 + \Delta_4)}{Z}  \\
\Gamma_2= \frac{\xi_2(\xi_1 + \Delta_1 + \xi_3 + \Delta_3 +\xi_4 + \Delta_4)}{Z}  \\
\Gamma_3=\frac{\xi_3(\xi_1 + \Delta_1 + \xi_2 + \Delta_2 +\xi_4 + \Delta_4)}{Z}  \\
\tiny{\Gamma_4= 1- (\Gamma_1 +\Gamma_2 + \Gamma_3)}
\end{gathered}
\end{equation}

\subsubsection{Selection}
The population for the next generation is selected by applying survival of the fittest concept using Eq. \ref{selection} where $\Phi_i^{j+1}$ is the selected candidate for next generation, $\chi_i^j $ is the solution generated after crossover, and $\Theta_i^j$ is currently existing candidate solution. 
\begin{gather}\label{selection}
\Phi_i^{j+1}=\begin{cases}
\chi_i^j & {If(fitness(\chi_i^j) \leq (fitness(\Theta_i^j))} \\
\Phi_i^j   & {\text{otherwise.}} 
\end{cases}
\end{gather}

\subsection{ Algorithm and Time complexity}
Algorithm \ref{euclid1} gives an operational summary of EQNN {optimization}. 
\begin{figure}[!htbp]
	\removelatexerror  
	\begin{algorithm}[H]
	\caption{EQNN training Algorithm}
	\label{euclid1}
			Initialize $CR_\mu = F_\mu =0.5 $, $CR_\sigma = 0.1$, $F_\sigma=0.3$, maximum generations $(M)$, $\Gamma_1$=$\Gamma_2$=$\Gamma_3$=$\Gamma_4$=0.25\; 
		
		Number of weight connection in each network  $D=(n+1)\times p + (p\times q) = p(n+q+1) \Rightarrow p(n+2)$ for q=1\;
		
		Initialize $N$ networks of length $D$ randomly\;
		
		Generate $CR_v$ and $M_v$ of size $N$, $CR_v=(CR_\mu, CR_\sigma,D)$, $F_v = (F_\mu, F_\sigma, N)$\;
		Evaluate each network on training data using fitness function\;
		
		\For{each generation $i \leq M$ } {
		Generate vector msp of $N$ random number $\in$ [0,1]\;
			\For{each network $j \leq N$} {
			{Generate $r_1 \ne r_2 \ne r_3 \ne r_4 \ne i \in [1,N]$ and $K_{rand} \in [1,D]$}\;
			\uIf {$0 \textless msp_i \leq  \Gamma_1$}			
			 { Apply$ DE/rand/1$ (Eq. \ref{eqn:mutation1})\;}
		
			\uElseIf {$ msp_i \leq (\Gamma_1 + \Gamma_2)$}{
			Apply $DE/best/1$ (Eq. \ref{eqn:mutation2})\;} 
		
			\uElseIf {$msp_i \leq (\Gamma_1 + \Gamma_2+ \Gamma_3)$}
			 { Apply $DE/current-to-best/1$ (Eq. \ref{eqn:mutation3})\;}		
		 
			\Else {Apply $DE/random/1$ (Eq. \ref{eqn:mutation4})\;}

			 Apply uniform crossover\;		
			}
		Evaluate newly generated solutions using error estimation function\;
		Select participants for next generation using Eq. \ref{selection}\;
		Update values of $\xi_1$, $\xi_2$, $\xi_3$, $\xi_4$, $\Delta_1$, $\Delta_2$, $\Delta_3$, $\Delta_4$\;
		 Update $\Gamma_1$, $\Gamma_2$, $\Gamma_3$, $\Gamma_4$ after $lp^M$ generations\;
		 Update ${CR}_v$ after $lp^{CR}$ generations\;    
	}
		
	\end{algorithm} 
\end{figure}
The time complexity of SB-ADE depends on the number of networks (N), number of network weight connections (D), number of input nodes (n), hidden nodes (p), where $p \simeq n$. Hence, the total time complexity for a maximum $M$ number of generations comes out to be $M \times O(n^2 \times N \times D) \Rightarrow O(n^2dNM)$.

\section{Performance Evaluation }

\subsection{Experimental Set-up}
The simulation experiments are conducted on a server machine assembled with two Intel\textsuperscript{\textregistered} Xeon\textsuperscript{\textregistered} Silver 4114 CPU with a 40 core processor and 2.20GHz clock speed. The computation machine is deployed with 64-bit Ubuntu 16.04 LTS, having 128 GB RAM. The proposed work is implemented in Python version 3.7 with the details listed in Table \ref{table:name1}.

\begin{table}[htbp]
	\centering
	
	\caption[Table caption text] {Experimental set-up parameters and their values.}  
	\label{table:name1}
	\begin{tabular}{|l|r|}
		\hline
		Parameter    & Value  \\
		\hline
		Input neural nodes ($n$)     & 7-28 \\
		Hidden layer nodes ($p$)        & 4-20       \\
		Output layer nodes($q$)      & 1          \\
		Maximum epochs($G_{max}$)     & 250        \\
		Size of training data & 75\%            \\
		Mutation learning period( $lp^M$)& 10\\
		Crossover rate learning period($lp^{CR}$) & 10\\
		Number of population  & 15\\
		\hline
	\end{tabular}
\end{table}
\subsection{ Data Sets }
Three different categories of benchmark datasets have been chosen for thorough analysis of the proposed work including Cluster workload from Google Cluster Data (GCD) \cite{Reiss2011} and PlanetLab Virtual Machines (VMs) traces \cite{beloglazov2012optimal}, Web servers workload collected from NASA and Saskatchewan web server request traces \cite{internet} and High Performance Computing (HPC) grid workloads belonging to Grid workload Archive \cite{Grid2019workloads}. GCD and Planet Lab (PL) workloads entail behavior of cloud applications for cluster and big data analytics (e.g., Hadoop) which gives resource usage traces collected over a period of 29 days in May, 2011. The CPU and memory utilization are extracted and aggregated over the period of first ten days for different prediction intervals like, 5 min, 10 min, ..., 60 min. The Planet Lab dataset contains 11,746 VMs traces which are measured at interval of five minutes and collected in March and April, 2011.
\par  NASA (NS) and Saskatchewan (SK.) HTTP traces contain information of Host, Time-stamp, HTTP request, HTTP reply and Bytes sent in the reply in ASCII files. For experimental analysis, we have extracted the Time-stamp values from NASA and Saskatchewan web server logs, which are aggregated into different prediction intervals, and the resultant values are normalized. Many scientific applications actively use cloud environment to assist diverse scientific computing and achieve high performance. The proposed EQNN prediction model is evaluated on three HPC grid workloads including AuverGrid (AG) \cite{AuverGrid2019workload}, NorduGrid (NG) \cite{NorduGrid2019workload}, and SHARCNet (SCN) \cite{SHARCNet2019workload}. Table \ref{table:(NASA)workloadcharacteristic} shows the statistical characteristics of the evaluated workloads. 
\begin{table}[!htbp] 	
	\caption[Table caption text] {Characteristics of evaluated workloads from Cluster, Web and HPC applications }  
	\label{table:(NASA)workloadcharacteristic} 	
	\centering
	\resizebox{9cm}{!}{
		\begin{tabular}{ | l | l | l | l  | l | l | }
			\hline
		\textbf{Category}	&\textbf{ Workload} &\textbf{ Duration} & \textbf{ Jobs}& \textbf{Mean(\%)} & \textbf{St.dev} \\
			\hline
			\hline
			\multirow{3}{*}{\textbf{Cluster }} &GCD-CPU & 10 days&2 M &21.84&13.62\\
			&GCD-memory  & 10 days& 2 M& 19.55&16.6 \\
			&PL-CPU& 10 days	& 1.5 M& 19.77&14.55\\
			\hline
			\hline
			\multirow{2}{*}{\textbf{Web }}&	NASA-HTTP & 31 days& 1.8 M& 2.95&1.72\\
			&{Saskatchewan} &7 months&2.5 M &  2.97&1.73 \\ 					
			\hline
			\hline
			\multirow{3}{*}{\textbf{HPC}} &AuverGrid & 365 days& 2.3 M &1.27E+9 &4.18E+4 \\	
			&NorduGrid&60 days&122K&9.56E+4  & 4.64E+5\\
				&SHARCNet&11 days&188K&1.14E+9  &4.89E+5 \\
				\hline
	\end{tabular}}
\end{table}

\subsection{Prediction Accuracy Analysis and Comparison}
\begin{figure*}[!htbp]
	\centering	
	\subfigure[VM 1 (Periodically variable load)]{\includegraphics[width=0.9\linewidth, scale=2]{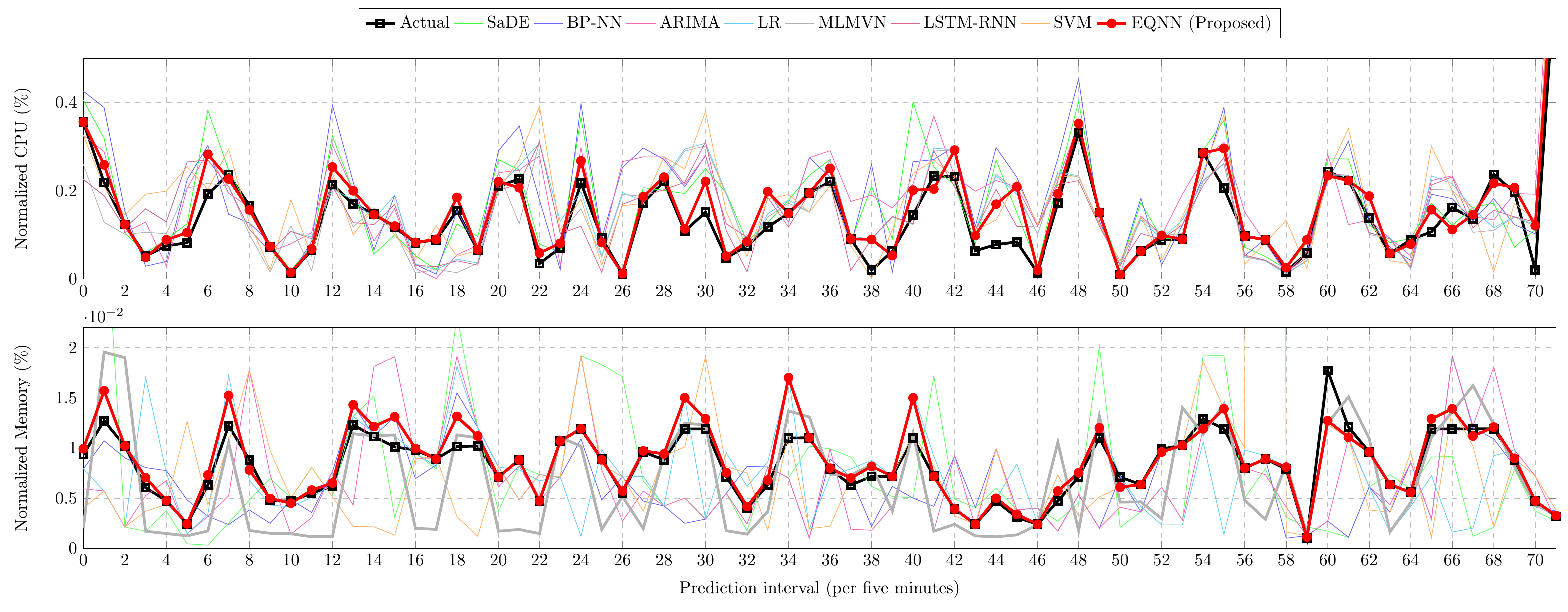}} 
	\subfigure[VM 2 (Randomly variable load)]{\includegraphics[width=0.9\linewidth, scale=2]{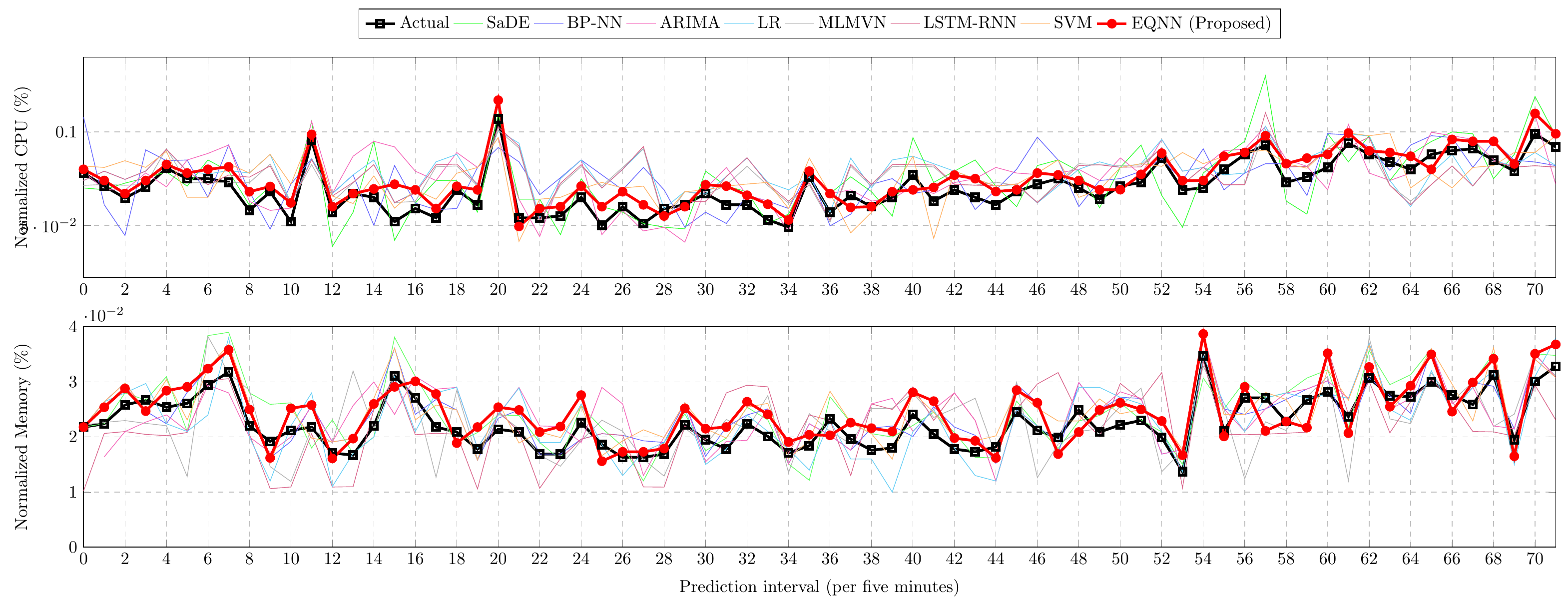}} 	
	\caption{Actual vs Predicted CPU and Memory utilization for two selected Google Cluster VMs}
	\label{VMs}	
\end{figure*}
The proposed EQNN prediction model is compared with widely used Support Vector Machine (SVM) \cite{singh2019tasm}, Linear Regression (LR) \cite{farahnakian2013lircup}, Autoregressive integrated moving average (ARIMA) \cite{baldan2016forecasting}, \cite{calheiros2014workload}, Neural Network (NN) based on Backpropagation (BPNN) \cite{lu2016rvlbpnn}, neural network optimized with Self-adaptive Differential Evolution (SaDE) algorithm \cite{kumar2018workload}, MLMVN \cite{qazi2018cloud}, LSTM-RNN \cite{kumar2018long} and Deep Learning \cite{zhang2018efficient} workload prediction models. The accuracy of EQNN prediction model is thoroughly investigated on a wide variety of cloud applications, variable workloads and estimation of resource utilization by VMs. Fig. \ref{VMs} compares actual and predicted resources (CPU and memory) utilization using EQNN and other comparative methods for a periodically and randomly variable workload by VM1 (Fig. \ref{VMs}(a)) and VM2 (Fig. \ref{VMs}(b)) respectively. The proposed method estimate the CPU and memory utilization closer to the actual resource utilization for both VMs serving a significantly different type of workload. Moreover, it is difficult to forecast in the presence of random peaks, as shown in the given Fig. \ref{VMs}, which are precisely estimated by the proposed method. In addition, the performance of the proposed work is analyzed for a different number of input and hidden neurons, where the most efficient outcome was obtained with ten input neurons as reported in Table \ref{table:inputhidden}. The number of hidden nodes were chosen to be seven \cite{kumar2018workload}. It has been noticed during experimental analysis that the performance of the model degrades as the number of input neurons increases beyond 10. The reason is that as the network size grows up, it starts memorizing the training set and not correlating or learning the patterns. On the other hand, with a smaller network size, it underestimates the useful information from the training set and shows a lesser ability to develop necessary correlations that result in performance degradation with reduced pattern learning. The subsequent sections entail the accuracy comparison for different workloads.

\begin{table}[!htbp]
	\centering
	\caption[Table caption text] {RMSE and Execution time elapsed for different combinations of Input nodes and Hidden nodes. }  
	\label{table:inputhidden}
	\resizebox{9cm}{!}{
		\begin{tabular}{|l|c|c|c|c|}
			\hline
			\textbf{Input nodes}& \textbf{Hidden nodes} & \textbf{RMSE} & \textbf{Training time (msec)} & \textbf{Iterations}  \\
			\hline
			\hline 
			7  & 4 & 0.0076 &116.67& 17 \\
			10 & 7 & 0.0022 & 234.89& 20\\
			15 & 10  & 0.0281 & 249.94& 22\\
			20 & 14 & 0.0430&  533.70& 19\\
			25 &  18 & 0.0615 & 618.11& 22\\
			\hline
	\end{tabular}}
\end{table}

\subsubsection{With Cluster workloads }
Fig \ref{fig:clusternormalized-rmse} shows the normalized RMSE results of predicted resource utilization by applying EQNN and other seven comparative approaches for the three cluster workloads, where $1.00$ indicates the result of proposed EQNN prediction model and higher values indicate worse performance. All the RMSE results are normalized with respect to RMSE output of EQNN. Also, the prediction error obtained by each method is quantified and visualized by computing absolute prediction error and analyzing its frequency for all three workloads. Fig. \ref{Clustererrorfrequency} shows the frequency of absolute error (\textit{Actual value} - \textit{Predicted value}), where EQNN yields minimum error for the majority of the workloads as compared to other seven methods. 
Table \ref{table:clusterMSE} presents RMSE value obtained for various prediction intervals including 5 min, 30 min and 60 minutes, where PWS is Prediction Window Size. The least error is obtained for 5 minutes prediction interval i.e., 0.00089 and 0.0018 for CPU and memory demand traces, respectively. It is found that RMSE increases with the size of prediction interval due to decrement in the number of data samples. On average, EQNN is 6\%-91.6\% more accurate than the seven comparative methods. This is because qubits have more diversity than real-valued weights and allow enhanced intuitive learning of the patterns during training process to precisely predict the unseen future variations.

\begin{figure*}[!htbp]
	\centering
		\resizebox{0.8\textwidth}{!}{\begin{tikzpicture}
	\begin{axis}[
	width=1.8\textwidth,
	height=0.45\textwidth,
	ybar,
	ymin=0,
	ymax=5,
	ybar=5pt,
	bar width=13pt,
	enlarge x limits=0.1,
	legend style={at={(0.5,.95)},
		anchor=north,legend columns=-1},
	ylabel={Normalized RMSE},
	symbolic x coords={G1, GCD-CPU, GCD-Memory,  PL-CPU,  Mean, G5},
	xtick=data,
	nodes near coords, nodes near coords align={vertical},
	ymajorgrids=true,
	xmajorgrids=true,
	grid style=dashed, thin
	]
	\addplot[draw=red,thick, fill=yellow!60!] coordinates{(GCD-CPU, 1) (GCD-Memory, 1) (PL-CPU, 1) (Mean, 1)}; \addlegendentry{EQNN}
	\addplot[fill=white, draw=green, pattern color=black!30, postaction={
		pattern=grid
	}] coordinates{(GCD-CPU, 1.88) (GCD-Memory, 2.27) (PL-CPU, 1.09) (Mean, 1.7566)}; \addlegendentry{SaDE-NN}
	\addplot+[fill=white,thick, pattern color=black!30, draw=blue, postaction={
		pattern=dots
	}] coordinates{(GCD-CPU, 1.4044) (GCD-Memory, 1.37) (PL-CPU, 3.52) (Mean, 2.097)}; \addlegendentry{BP-NN}
	\addplot+[fill=white, thick, pattern color=black!30, draw=magenta, postaction={
		pattern=vertical lines
	}] coordinates{(GCD-CPU, 1.25) (GCD-Memory, 2.84) (PL-CPU, 4.03) (Mean, 2.70262)};\addlegendentry{ARIMA} 
	\addplot[fill=white,thick, draw=cyan,pattern color=black!30, postaction={
		pattern=north east lines
	}] coordinates{(GCD-CPU, 1.30) (GCD-Memory, 3.34) (PL-CPU, 3.45) (Mean, 2.69)}; \addlegendentry{LR}
	\addplot[fill=white, thick, pattern color=black!30, draw=orange!80!,  postaction={
		pattern=horizontal lines
	}] coordinates{(GCD-CPU, 1.59) (GCD-Memory, 3.43) (PL-CPU, 3.98) (Mean, 3.0)}; \addlegendentry{SVM}
	\addplot[fill=white, thick, pattern color=black!20, draw=black!70!,  postaction={
	pattern=north west lines
}] coordinates{(GCD-CPU, 2.21) (GCD-Memory, 2.00) (PL-CPU, 1.41) (Mean, 1.87)}; \addlegendentry{MLMVN}
	\addplot[fill=white, thick, pattern color=black!20, draw=purple,  postaction={
	pattern=crosshatch
}] coordinates{(GCD-CPU, 1.38) (GCD-Memory, 1.07) (PL-CPU, 1.473) (Mean, 1.31)}; \addlegendentry{LSTM-RNN}

	\addplot[red,line legend,sharp plot,nodes near coords={},
	update limits=false,shorten >=-3mm,shorten <=-3mm] 
	coordinates {(G1, 1) (G5, 1)} 
	node[midway,above,font=\bfseries\sffamily]{EQNN (1.00)};
	\end{axis}
	\end{tikzpicture}}
	\caption{Normalized RMSE results in cluster workloads }
	\label{fig:clusternormalized-rmse}
\end{figure*}
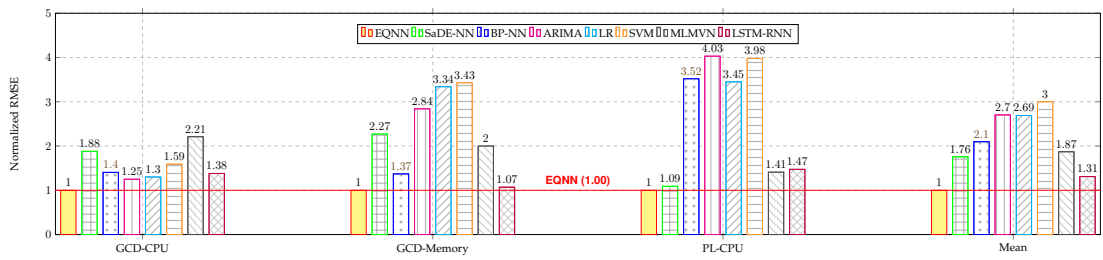

\begin{figure*}[!htbp]
	\centering	
	\subfigure[GCD-CPU]{\includegraphics[width=0.31\linewidth, scale=2]{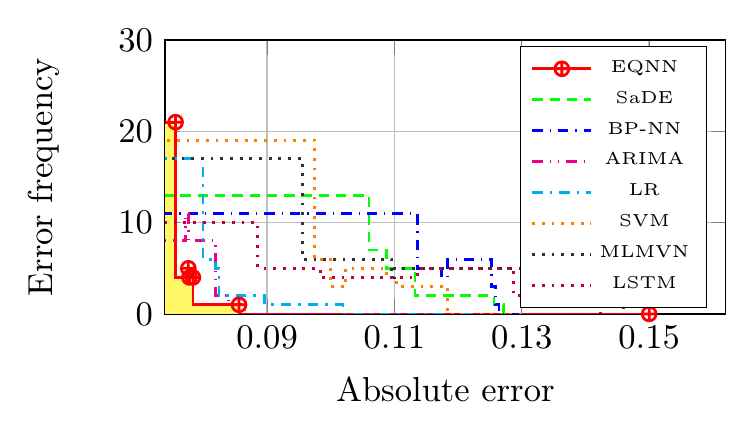}} 
	\subfigure[GCD-Memory]{\includegraphics[width=0.31\linewidth, scale=2]{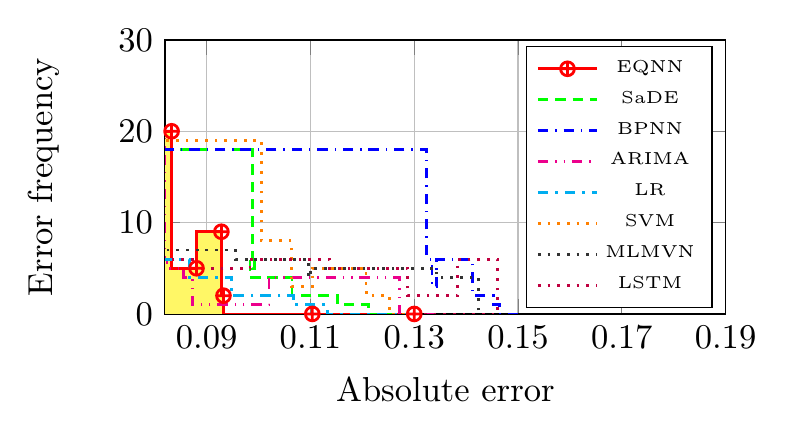}} 
	\subfigure[PL-CPU]{\includegraphics[width=0.31\linewidth, scale=2]{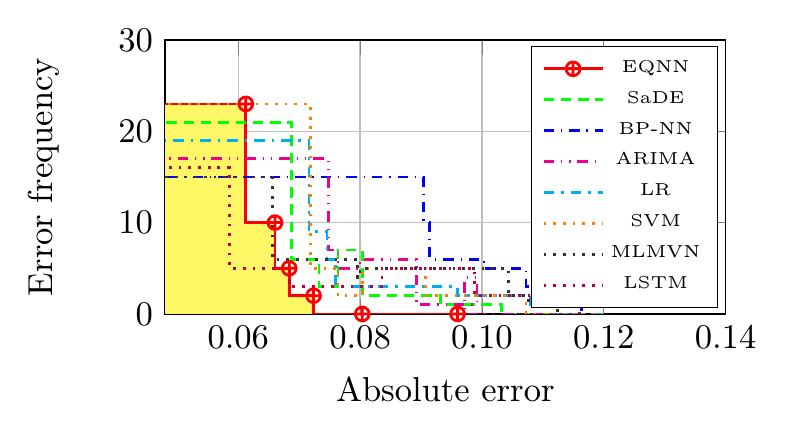}} 	
	\caption{Absolute error frequency of resource utilization estimation for cluster workloads}
	\label{Clustererrorfrequency}
	
\end{figure*}

\begin{table}[!htbp]
	\centering
	\caption[Table caption text] {RMSE ($10^{-3}$) for Cluster workloads (DS:Datasets) }  
	\label{table:clusterMSE}
	\resizebox{9cm}{!}{
		\begin{tabular}{|l|c|c|c|c|c|c|c|c|c|}
			\hline			
			\textbf{DS}& \textbf{PWS}   & \textbf{SVM}  & \textbf{LR} & \textbf{ARIMA} & \textbf{BP} &\textbf{SaDE}& \textbf{MLM}&\textbf{LSTM}& \textbf{EQNN} \\
			 & (min)& & & & & &\textbf{-VN} &\textbf{-RNN} & \\ 	\hline \hline
			GCD-CPU& \multirow{3}{*}{ 5}& 1.42&1.16 &1.11 & 1.25&1.68 &1.97 &1.23 &0.89 \\
			GCD-Mem& &6.18 &6.01 &5.12 &2.48 &4.08 &3.61 &1.91 &1.80 \\ 
			PL-CPU& &8.41 &7.30 &8.51 &7.43 &2.30&2.98 & 3.10 &2.11 \\ \hline
			GCD-CPU	& \multirow{3}{*}{ 30}&9.40 & 8.84&11.9 & 27.9&12.3 &4.24 &1.30 &1.14 \\
			GCD-Mem& &9.41 & 8.85&1.70 & 20.8&6.78&6.21 & 9.13 &1.33 \\ 
			PL-CPU& &16.6 &15.3 &12.3 & 25.6& 15.3&53.1 &20.4 &14.6\\ \hline
			GCD-CPU& \multirow{3}{*}{ 60 }&9.46 &10.17 &11.7 & 17.9&34.9 &14.7 &10.2 &2.92 \\ 
			GCD-Mem& &10.5 &12.02 &13.50 & 30.6&8.92&12.6 &11.3  &9.60 \\ 
			PL-CPU& &6.14 &5.36 &4.70 &20.8 &10.6&12.2 & 18.7 &9.21 \\ \hline
			
	\end{tabular}}
\end{table}

\subsubsection{With Web workloads}
Fig \ref{fig:webnormalized-rmse} shows the normalized RMSE bar plots of actual and predicted workload for job arrival during two prediction intervals viz. 5 minutes and 60 minutes using EQNN and other seven comparative approaches for NASA and Saskatchewan HTTP traces. The resulted plots clearly depict that EQNN improves prediction accuracy by more than 90\% against all the seven methods for each prediction interval in case of web workloads. The frequency of absolute error for prediction results of both NASA and Saskatchewan HTTP traces are plotted in Fig. \ref{Errorfreq_nasa_sask}, where proposed EQNN always delivers the least prediction error as compared to other comparative approaches. Moreover, the high frequency of absolute error in the case of EQNN signifies the robustness and strong tendency for yielding maximum prediction accuracy. 
\begin{figure*}[!htbp]
	\centering
	\resizebox{0.8\textwidth}{!}{	\begin{tikzpicture}
	\begin{axis}[
	width=1.8\textwidth,
	height=0.5\textwidth,
	ybar,
	ymin=0,
	ymax=13,
	ybar=5pt,
	bar width=13pt,
	enlarge x limits=0.1,
	legend style={at={(0.5,.95)},
		anchor=north,legend columns=-1},
	ylabel={Normalized RMSE},
	ymajorgrids=true,
	xmajorgrids=true,
	grid style=dashed, thin,
	symbolic x coords={G1, NASA (5 min), NASA (60 min),  SASK. (5 min),   SASK. (60 min), Mean, G5},
	xtick=data,
	nodes near coords, nodes near coords align={vertical}]
	\addplot[draw=red, thick, fill=yellow!60!] coordinates{(NASA (5 min), 1) (NASA (60 min), 1) (SASK. (5 min), 1)  (SASK. (60 min), 1) (Mean, 1)}; \addlegendentry{EQNN}
	\addplot[fill=white, draw=green, pattern color=black!30, thick, postaction={
		pattern=grid
	}] coordinates{(NASA (5 min),1.4) (NASA (60 min), 5.68) (SASK. (5 min), 1.81)  (SASK. (60 min), 5.4) (Mean, 3.57)}; \addlegendentry{SaDE-NN}
	\addplot+[fill=white, draw=blue, pattern color=black!30, thick, postaction={
		pattern=dots
	}] coordinates{(NASA (5 min), 4.9) (NASA (60 min), 12.2) (SASK. (5 min),8.5 )  (SASK. (60 min), 10.4) (Mean, 9.0)}; \addlegendentry{BP-NN}
	\addplot+[fill=white, draw=magenta,pattern color=black!30, thick, postaction={
		pattern=vertical lines
	}] coordinates{(NASA (5 min), 6.64) (NASA (60 min), 2.18) (SASK. (5 min), 2.5)  (SASK. (60 min), 2.14) (Mean, 3.4)};\addlegendentry{ARIMA} 
	\addplot[fill=white, draw=cyan, pattern color=black!30,thick, postaction={
		pattern=north east lines
	}] coordinates{(NASA (5 min), 6.21) (NASA (60 min), 2.12) (SASK. (5 min), 2.54)  (SASK. (60 min), 2.16) (Mean, 3.3)}; \addlegendentry{LR}
	\addplot[fill=white,draw=orange!80!, pattern color=black!30, thick, postaction={
		pattern=horizontal lines
	}] coordinates{(NASA (5 min), 6.9) (NASA (60 min), 2.86) (SASK. (5 min), 3.21)  (SASK. (60 min), 2.06) (Mean, 3.8)}; \addlegendentry{SVM}

	\addplot[fill=white, thick, pattern color=black!20, draw=black!70!,  postaction={
	pattern=north west lines
}] coordinates{(NASA (5 min), 2.94) (NASA (60 min), 3.96) (SASK. (5 min), 6.54) (SASK. (60 min), 1.06) (Mean, 3.62)}; \addlegendentry{MLMVN}
\addplot[fill=white, thick, pattern color=black!20, draw=purple,  postaction={
	pattern=crosshatch
}] coordinates{(NASA (5 min), 1.54) (NASA (60 min), 2.24) (SASK. (5 min), 2.88) (SASK. (60 min), 1.45) (Mean, 1.45)}; \addlegendentry{LSTM-RNN}
	\addplot[red,line legend,sharp plot,nodes near coords={},
	update limits=false,shorten >=-3mm,shorten <=-3mm] 
	coordinates {(G1, 1) (G5, 1)} 
	node[pos=0.415 ,above,font=\bfseries\sffamily]{EQNN };
	\addplot[red,line legend,sharp plot,nodes near coords={},
	update limits=false,shorten >=-3mm,shorten <=-3mm] 
	coordinates {(G1, 1) (G5, 1)} 
	node[pos=0.415 ,below,font=\bfseries\sffamily]{(1.00)};
	\end{axis}
	\end{tikzpicture}}
	\caption{Normalized RMSE results in Web workloads}
	\label{fig:webnormalized-rmse}
\end{figure*}
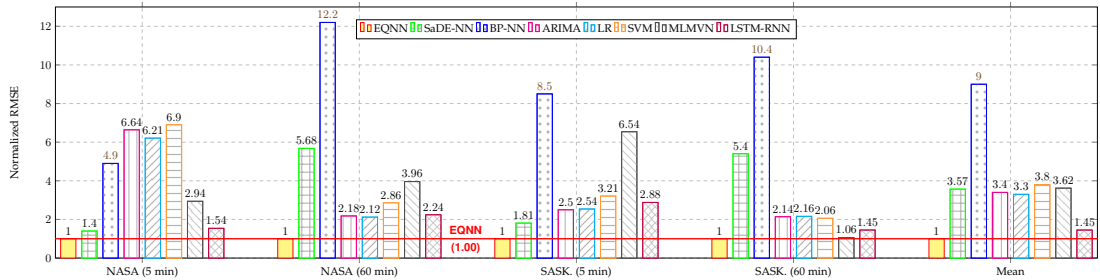
\begin{figure}[!htbp]
	\centering	
	\subfigure[NASA HTTP traces]{\includegraphics[width=0.24\textwidth, scale=2]{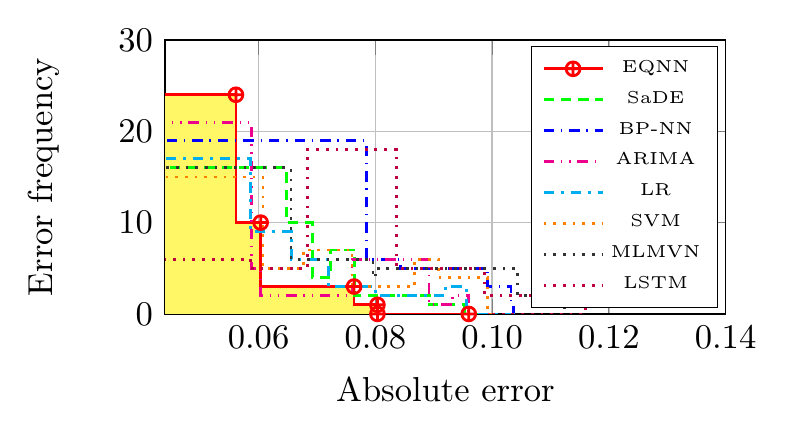}} 
	\subfigure[Saskatchewan  HTTP traces ]{\includegraphics[width=0.24\textwidth, scale=2]{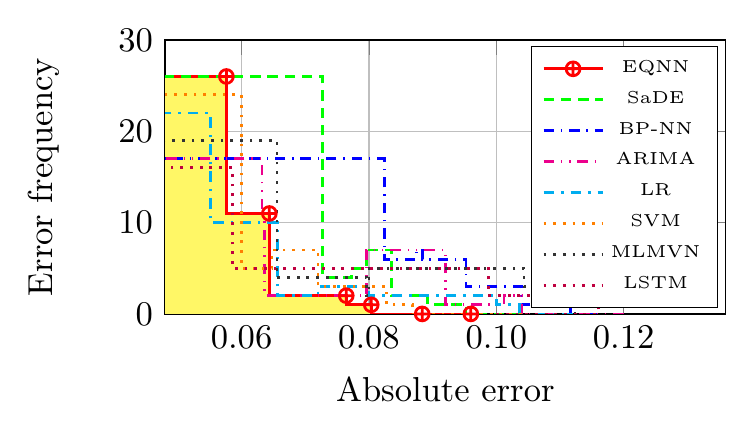}} 
	
	\caption{Absolute error frequency}
	\label{Errorfreq_nasa_sask}
	
\end{figure}
Table \ref{table:webMSE} shows comparative prediction accuracy of existing and proposed EQNN prediction model over different web workloads for different prediction intervals, where least prediction error is obtained with Prediction Window Size (PWS) of five minutes, and thereafter prediction error increases slightly due to decrease in the number of training data samples for bigger PWS.
\begin{table}[!htbp]
	\centering
	\caption[Table caption text] { RMSE ($10^{-3}$) for Web workloads}  
	\label{table:webMSE}
	\resizebox{9cm}{!}{
		\begin{tabular}{|l|c|c|c|c|c|c|c|c|c|}
		\hline			
		\textbf{DS}& \textbf{PWS}   & \textbf{SVM}  & \textbf{LR} & \textbf{ARIMA} & \textbf{BP} &\textbf{SaDE}& \textbf{MLM}&\textbf{LSTM}& \textbf{EQNN} \\
		&\textbf{ (min)}& & & & & & \textbf{-VN}&\textbf{-RNN} & \\ 	\hline \hline
			NS& \multirow{2}{*}{ 5}&4.92 &4.41 &4.72 &3.51 &1.00&{2.09}
			&1.10 &0.71 \\
			SK.& &3.54 &2.86 &2.75 &9.36 &2.00& 7.2&{3.17} &1.10 \\ \hline
			NS& \multirow{2}{*}{ 30}&5.97 & 4.71&4.30 &29.2 &14.2& 6.91&{6.43} &2.50 \\
			SK.& &4.82 &4.53 &4.31 &34.9 &20.9&26.5 &{4.79}&3.50 \\ \hline
			NS& \multirow{2}{*}{ 60}&7.15 &5.31 &5.47 & 30.5&14.2& 9.91&{5.59} & 2.50\\
			SK.& &6.40 &6.70 &6.66 &32.3 &17.0& 3.31&{4.50}&3.10 \\ \hline

			\hline	
	\end{tabular}}
\end{table}
The reason for such an improved accuracy of predicted values is thorough learning of useful information by QNN model from input data samples. The C-NOT gate allows a rotational effect that traces the input data intensively and generates counter-intuitive patterns from it, and closely anticipates the future workload information. 
\subsubsection{With HPC Workloads}
The normalized RMSE results of predicted resource utilization by applying EQNN and other seven comparative approaches for the three scientific or high performance computing workloads are shown in Fig \ref{fig:hpcnormalizedrmse}. On average, EQNN achieves prediction accuracy improvement of 25\% - 96\% over the seven comparative methods. Fig. \ref{HPCerrorfrequency} shows the frequency of absolute error, where EQNN yield minimum error for the majority of the workloads as compared to other seven methods. 
Table \ref{table:HPCMSE} presents RMSE value obtained for various prediction intervals including 5 min, 30 min, and 60 minutes. The forecasting accuracy is improved due to the usage of qubits in the form of phase values which draws counter intuitive patterns based on behavior of the application and allows extensive exploration and better learning by neural network as compared to real valued neural network. Moreover, the application of an optimally adaptive differential evolution algorithm for training of EQNN prediction model with effective balancing of exploration and exploitation has enhanced the learning capabilities of prediction model. It allows searching in multiple directions and applies a greedy approach as well as raises diversity of the search space to find out optimal/near-optimal solution. Additionally, the proposed work is compared with prediction results achieved in ARIMA and SVM \cite{baldan2016forecasting} on google cluster CPU traces by applying Mean Absolute Error (MAE) for PWS of 5 and 60 minutes as shown in Table \ref{table:gcdcomparison}, where the $-$ represents unavailability of the corresponding result. The RMSE results of EQNN and Deep learning  \cite{zhang2018efficient} obtained on Planet Lab CPU traces are also compared in Table \ref{table:PLcomparison}. The comparison of training-time of the proposed and comparative approaches is shown in Table \ref{table:timecomparison}, which is highest for ARIMA, followed by the proposed approach (due to generation of qubit-based network weights) and least for LR. However, the efficiency and applicability of the proposed predictor is not affected because the training is a periodic task and can be executed in parallel on the servers equipped with enough resources.
\begin{figure*}
	\centering
\resizebox{0.8\textwidth}{!}{	
	\begin{tikzpicture}
	\begin{axis}[
	width=1.5\textwidth,
	height=0.5\textwidth,
	ybar,
	ymin=0,
	ymax=5.5,
	ybar=4pt,
	ymajorgrids=true,
	xmajorgrids=true,
	grid style=dashed, thin,
	bar width=13pt,
	enlarge x limits=0.15,
	legend style={at={(0.5,.95)},
		anchor=north,legend columns=-1},
	ylabel={Normalized RMSE},
	symbolic x coords={G1, AuverGrid, NorduGrid, ShareCNet, Mean, G5},
	xtick=data,
	nodes near coords, nodes near coords align={vertical}]
	\addplot[draw=red, thick, fill=yellow!60!] coordinates{(AuverGrid, 1) (NorduGrid, 1) (ShareCNet,1)  (Mean, 1)}; \addlegendentry{EQNN}
	\addplot[fill=white, draw=green, pattern color=black!30, thick, postaction={
		pattern=grid
	}] coordinates{(AuverGrid, 1.065) (NorduGrid, 1.81) (ShareCNet,1.76) (Mean, 1.54375)}; \addlegendentry{SaDE-NN}
	\addplot+[fill=white, draw=blue, thick,
	pattern=dots,
	pattern color=black!30,  
	] coordinates{(AuverGrid, 1.794) (NorduGrid, 2.25) (ShareCNet,4.55) (Mean, 2.864)}; \addlegendentry{BP-NN}
	\addplot+[fill=white, draw=magenta, thick, postaction={
		pattern=vertical lines, pattern color=black!30,
	}] coordinates{(AuverGrid, 1.93) (NorduGrid, 1.1737) (ShareCNet,1.16)  (Mean, 1.4152)};\addlegendentry{ARIMA} 
	\addplot[fill=white, draw=cyan, thick,pattern color=black!30,  postaction={
		pattern=north east lines, dashed, 
	}] coordinates{(AuverGrid, 1.4006) (NorduGrid, 1.107) (ShareCNet,3.93)  (Mean, 2.1453)}; \addlegendentry{LR}
	\addplot[fill=white,draw=orange!80!, thick, postaction={
		pattern=horizontal lines, pattern color=black!30,
	}] coordinates{(AuverGrid, 1.47) (NorduGrid, 1.122) (ShareCNet,3.70)  (Mean, 2.097)}; \addlegendentry{SVM}

	\addplot[fill=white, thick, pattern color=black!20, draw=black!70!,  postaction={
	pattern=north west lines
}] coordinates{(AuverGrid, 1.79) (NorduGrid, 1.93) (ShareCNet, 1.86) (Mean, 2.4)}; \addlegendentry{MLMVN}
\addplot[fill=white, thick, pattern color=black!20, draw=purple,  postaction={
	pattern=crosshatch
}] coordinates{(AuverGrid, 1.16) (NorduGrid, 1.17) (ShareCNet, 2.05) (Mean, 1.46197)}; \addlegendentry{LSTM-RNN}
	\addplot[red,line legend,sharp plot,nodes near coords={},
	update limits=false,shorten >=-3mm,shorten <=-3mm] 
	coordinates {(G1, 1) (G5, 1)} 
	node[pos=0.5 ,above,font=\bfseries\sffamily]{EQNN };
	\addplot[red,line legend,sharp plot,nodes near coords={},
	update limits=false,shorten >=-5mm,shorten <=-5mm] 
	coordinates {(G1, 1) (G5, 1)} 
	node[pos=0.5 ,below,font=\bfseries\sffamily]{(1.00)};
	\end{axis}
	\end{tikzpicture}}
	\caption{Normalized RMSE results in HPC workloads}
	\label{fig:hpcnormalizedrmse}
\end{figure*}
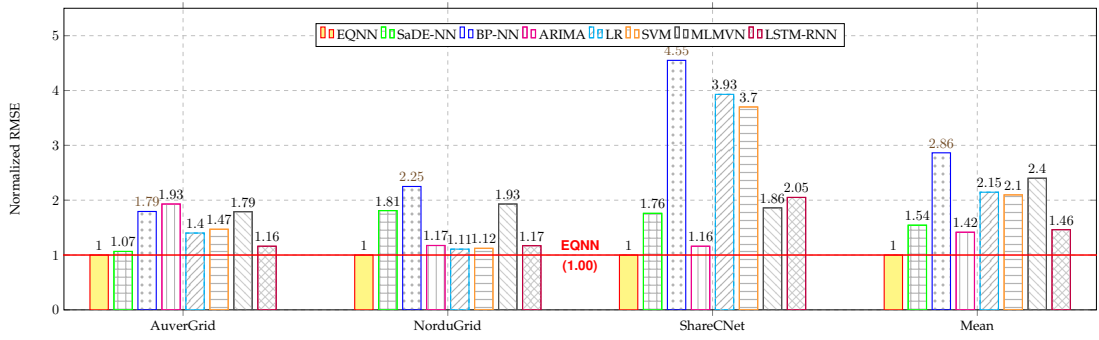

\begin{figure*}[!htbp]
	\centering	
	\subfigure[AuverGrid]{\includegraphics[width=0.31\linewidth, scale=2]{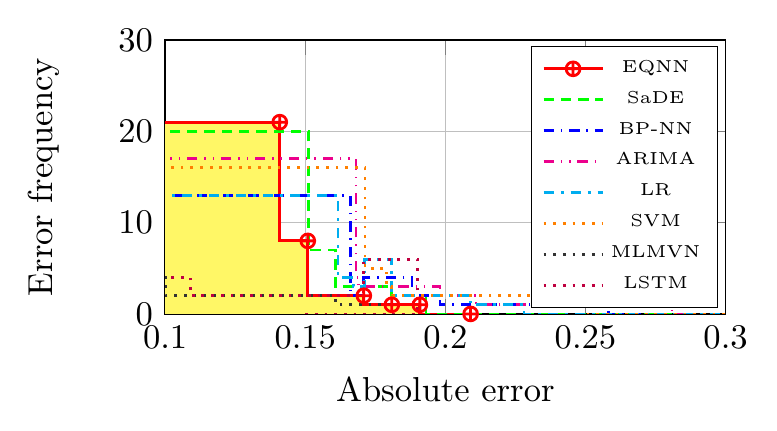}} 
	\subfigure[NorduGrid]{\includegraphics[width=0.31\linewidth, scale=2]{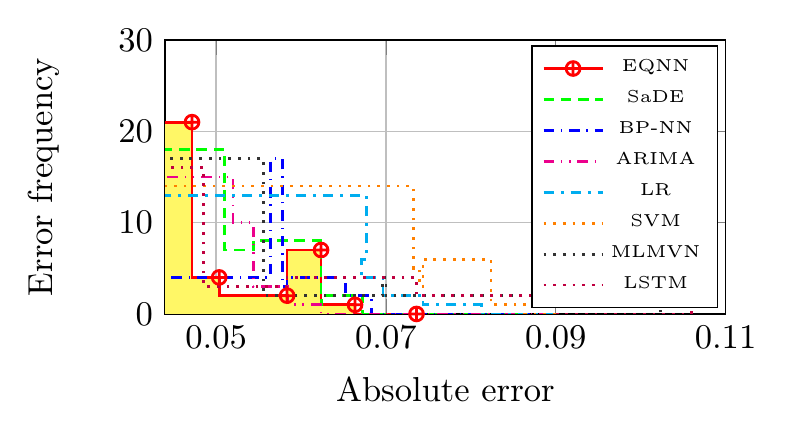}} 
	\subfigure[ShareCNet]{\includegraphics[width=0.31\linewidth, scale=2]{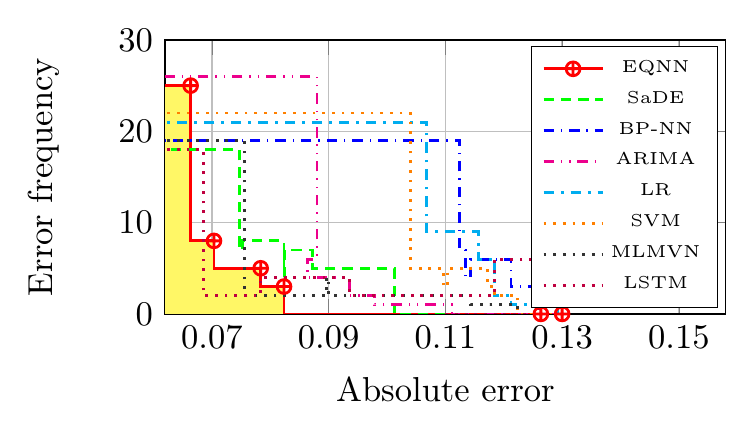}} 	
	\caption{Absolute error frequency of number of jobs estimation for HPC workloads}
	\label{HPCerrorfrequency}
	
\end{figure*}
\begin{table}[!htbp]
	\centering
	\caption[Table caption text] { RMSE ($10^{-3}$) over Scientific/HPC Grid workloads}  
	\label{table:HPCMSE}
	\resizebox{9cm}{!}{
		\begin{tabular}{|l|c|c|c|c|c|c|c|c|c|}
			\hline			
			\textbf{DS}& \textbf{PWS}   & \textbf{SVM}  & \textbf{LR} & \textbf{ARIMA} & \textbf{BP} &\textbf{SaDE}& \textbf{MLM}&\textbf{LSTM}& \textbf{EQNN} \\
			&\textbf{ (min)}& & & & & &\textbf{-VN} &\textbf{-RNN} & \\ 	\hline \hline
		
			AuverG.& \multirow{3}{*}{ 5 min}& 4.45&4.23 &5.83 &5.42 &3.21&5.42 &3.52 &3.02 \\
			NorduG.& &2.39 &2.36 &2.50 &4.80 &3.89 &4.12 & 3.05&2.13 \\ 
			SHARCNet& &9.6 & 10.2&3.00 &11.8 &4.56 &4.82 &5.3 &2.59 \\ \hline
			AuverG.& \multirow{3}{*}{ 30 min}&7.78 &9.38 &9.10 &9.52 &8.92 &12.2 &8.95 &6.71 \\
			NorduG.& &32.5 &28.5 &43.5 &48.2 &13.9 &18.9 & 32.2&10.2 \\ 
			SHARCNet& &5.78 &4.10 &20.0 &6.76 &8.93 &11.8 &9.92 &4.79 \\ \hline
			AuverG.& \multirow{3}{*}{ 60 min}& 269.6&272.6 &22.4 &17.8 &20.1 &9.2 &14.0 &10.6 \\
			NorduG.& &13.8 &14.7 & 19.2&22.9 &18.3 &25.6 & 17.9&11.2 \\ 
			SHARCNet& &25.2 &33.9 &10.6 &25.4 & 12.2&22.4 & 14.4& 6.35\\ \hline

	\end{tabular}}
\end{table}

\begin{table}[!htbp]
	\centering
	\caption[Table caption text] {MAE comparison on GCD-CPU dataset }  
	\label{table:gcdcomparison}
		\begin{tabular}{|l|c|c|c|}
			\hline			
			Approach&   	ARIMA \cite{baldan2016forecasting}   & SVM \cite{baldan2016forecasting} & 	Proposed \\ 	\hline
		5 min &0.9720&0.9690 & 0.3715 \\ 	
		60 min	&0.9210&- & 0.4827 \\ 	
		
			\hline	
	\end{tabular}
\end{table}
\begin{table}[!htbp]
	\centering
	\caption[Table caption text] {RMSE ($10^{-3}$) comparison on Planet Lab CPU}  
	\label{table:PLcomparison}
		\begin{tabular}{|l|c|c|c|}
			\hline			
			Approach&	5 min    & 30 min  & 60 min   \\ 	\hline
			Deep learning  \cite{zhang2018efficient} &9.1700&10.3000 & 9.9700 \\ 	
		
			Proposed& 0.0021 & 0.0146 & 0.00921 \\
			
			\hline	
	\end{tabular}
\end{table}

\begin{table}[!htbp]
	\centering
	\caption[Table caption text] {{Training-time (sec) for prediction-intervals}}  
	\label{table:timecomparison}
	\begin{tabular}{|l|c|c|c|}
		\hline			
		{\textbf{Approach}}&{	5 min }   & {30 min } & {60 min}   \\ 	\hline \hline
	{\textbf{SVM}}&{1.08}&{0.50} & {0.25} \\ 	
	{\textbf{LR}}	 &{0.84 }&{0.38 } &{ 0.24}\\ 	
	{\textbf{BP}} &{24.11} &{ 4.27} & {2.91}\\ 	
	{\textbf{SaDE}} & {48.58}& { 4.27}&{2.91}\\ 
	{\textbf{ARIMA}} &{1769.07} &{75.18}  & {29.05}\\ 	
		{\textbf{MLMVN}} &{179.11} &{36.10}  & {19.23}\\ 	
			{\textbf{LSTM-RNN}} &{193.07} &{39.28}  & {16.45}\\ 			
	{	\textbf{Proposed (EQNN)}}& {223.41} & {26.01} &{ 11.91} \\
		
		\hline	
	\end{tabular}
\end{table}

\subsection{Statistical Analysis}
The obtained results are validated by conducting a statistical test on STAC \cite{rodriguez2015stac} web platform and applying the Friedman test followed by Finner post hoc analysis. The Friedman test considers a null hypothesis ($H_0$) which assumes that there is no significant difference in the results of different approaches. Finner post hoc analysis analyzes the pairwise performance of the algorithms. The tests are conducted by using SB-ADE algorithm as a control method with a significance level of 0.05. Table \ref{table:friedman} shows that the Friedman test rejects the $H_0$ for NASA, Saskatchewan HTTP traces, Planet Lab CPU and NorduGrid workload indicates the presence of a significant difference in the results. However, it is accepted for Google Cluster CPU and memory; AuverGrid and SHARCNet specify the absence of this difference. 
The ranks obtained by different approaches through the Friedman test are shown in Table \ref{table:rank}. It can be observed that the best rank is obtained by the proposed method in majority of the cases. The Finner analysis given in Table \ref{table:finner} shows a pair-wise comparison of presence of significant difference between the two approaches where the comparison of the pairs are rejected and accepted as well which indicates the presence of a significant difference and no difference between them, respectively. Hence, the proposed EQNN based prediction approach outperforms the state-of-the-art prediction methods, and it is suitable to enhance the accuracy of workload forecasting at the cloud datacenter.
\begin{table}[htbp]
	\centering
	\caption[Table caption text] {Friedman Test Result }  
	\label{table:friedman}
	\resizebox{8cm}{!}{
		\begin{tabular}{|l|c|c|c|}
			\hline
			\textbf{Dataset} & \textbf{Statistic} & \textbf{p-value} & \textbf{Result}
			  \\ 	\hline \hline
			GCD-CPU &2.92 & 0.07009&$H_0$ is accepted \\
			GCD-Mem. &2.26 & 0.12803&$H_0$ is accepted \\
			PL-CPU &12.00 & 0.000058&$H_0$ is rejected \\
			\hline  
			NASA HTTP&822.27 & 0.000055&$H_0$ is rejected \\
			Saskatchewan HTTP& 322.27 & 0.00069&$H_0$ is rejected\\
						\hline	
		    AuverGrid&2.92 & 0.07009&$H_0$ is accepted \\
		    NorduGrid&11.13 & 0.000079&$H_0$ is rejected \\
		    SHARCNet&2.37 & 0.11452&$H_0$ is accepted \\
		    \hline
	\end{tabular}}
\end{table}
\begin{table*}[htbp]
	\centering
	\caption[Table caption text] {Friedman Test Ranks }  
	\label{table:rank}
		\begin{tabular}{|l|c|c|c|c|c|c|c|c|}
			\hline
			\hline
			Approach &GCD-CPU&GCD-Mem.&PL-CPU& NASA & Saskatchewan & AuverGrig& NorduGrid&SHARCNet
			  \\ 	\hline \hline
		EQNN& 1.6667&1.3333&3.6667 &1.0000&1.0833&1.0000&1.0000&1.3333 \\
		SaDE&5.3333 &2.3333&4.6667 &2.9166&2.5833&2.6667&3.6667&3.6667 \\
		ARIMA& 3.3333 &4.0000&1.3333 &2.0833&2.3333&4.6667&4.6667&4.3333 \\
     	LR & 2.33333 &4.0000&1.6667 &3.6667&3.3333&4.6667&2.6667&2.3333 \\
			SVM & 3.3333 &4.6667&3.8333 &3.3333 &3.6667&3.6667 &3.0000&4.6667 \\
			BP-NN& 5.0000 &4.6667&5.8333 &4.0000&4.166&4.3333&6.0000&4.6667 \\
			\hline	
	\end{tabular}
\end{table*}

\begin{table*}[htbp]
	\centering
	\caption[Table caption text]{Finner post-hoc analysis result }
	\label{table:finner}
	
		\begin{tabular}{|l|c|c|c||c|c|c|}  
			\hline
			\textbf{Comparison}	& \multicolumn{3}{c}{\textbf{NASA HTTP} }&\multicolumn{3}{c|}{\textbf{Saskatchewan HTTP}}\\ 
			\hline
			& Statistics & Adjusted p-value& $H_0$ &  Statistics & Adjusted p-value& $H_0$\\
				\cline{2-4} \cline{5-7} 
			EQNN vs SVM& 4.38&0.00005&rejected& 4.10&0.00016&rejected\\
			EQNN vs LR & 2.27 & 0.0060&rejected& 2.37&0.0014&rejected\\
			EQNN vs ARIMA& 1.18&0.23533& accepted & 2.16&0.03038 & rejected\\
			EQNN vs BPNN & 3.28 & 0.0020&rejected  & 3.37& 0.0014&rejected \\
			EQNN vs SaDE & 2.09 & 0.0047& rejected & 2.59& 0.01866&rejected \\
			\hline \hline 
				\textbf{Comparison}	&  \multicolumn{3}{c}{\textbf{GCD-CPU} }&\multicolumn{3}{c|}{\textbf{PL-CPU}}\\ 
				\hline
			& Statistics & Adjusted p-value& $H_0$ &  Statistics & Adjusted p-value& $H_0$\\ 		\cline{2-4} \cline{5-7}  
				EQNN vs SVM& 1.09&0.4152&accepted& 0.109&0.91312&accepted\\
			EQNN vs LR & 0.44 & 0.4364&accepted& 1.3093&0.49185&accepted\\
			EQNN vs ARIMA& 1.09&0.4152&accepted & 1.52759&0.49185 & accepted\\
			EQNN vs BPNN & 2.18 & 0.0792& accepted & 1.4184& 0.49185&accepted\\
			EQNN vs SaDE & 2.40 & 0.0792& accepted & 0.65465& 0.59285&accepted \\
			\hline \hline 
				\textbf{Comparison}	&  \multicolumn{3}{c}{\textbf{AuverGrid} }&\multicolumn{3}{c|}{\textbf{NorduGrid}}\\ 
				\hline
			&Statistics & Adjusted p-value& $H_0$ &  Statistics & Adjusted p-value& $H_0$\\ 			\cline{2-4} \cline{5-7} 
		EQNN vs SVM& 1.75&0.10003&accepted&1.30&0.23208&accepted\\
			EQNN vs LR & 2.40 & 0.07925&accepted& 1.09&0.27523&accepted\\
			EQNN vs ARIMA& 2.40&0.07925& accepted & 2.40&0.04044 & rejected\\
			EQNN vs BPNN & 2.18 & 0.07925& accepted & 3.27& 0.00530&rejected \\
			EQNN vs SaDE & 1.09 & 0.27523& accepted & 1.74& 0.13109&rejected \\
			\hline 
	\end{tabular}

\end{table*}

\section{ Conclusions and Future work} 
This work proposed a novel EQNN model to forecast future workload demands at the cloud datacenter. The input data points and neural weight connections are qubit phase values. A self-balanced adaptive differential evolution (SB-ADE) algorithm is developed to train this model. The experimental analysis of the proposed work is done by using eight benchmark real workload datasets, and the EQNN model surpasses the prediction accuracy over state-of-the-art approaches. The future aim of this work would be to raise the quality of prediction further and allow online training with live data for prediction of bandwidth requirement and power consumption in future at cloud datacenters.

\bibliographystyle{IEEEtran}
\bibliography{bibfile}
\begin{IEEEbiography}[{\includegraphics[width=0.8\linewidth]{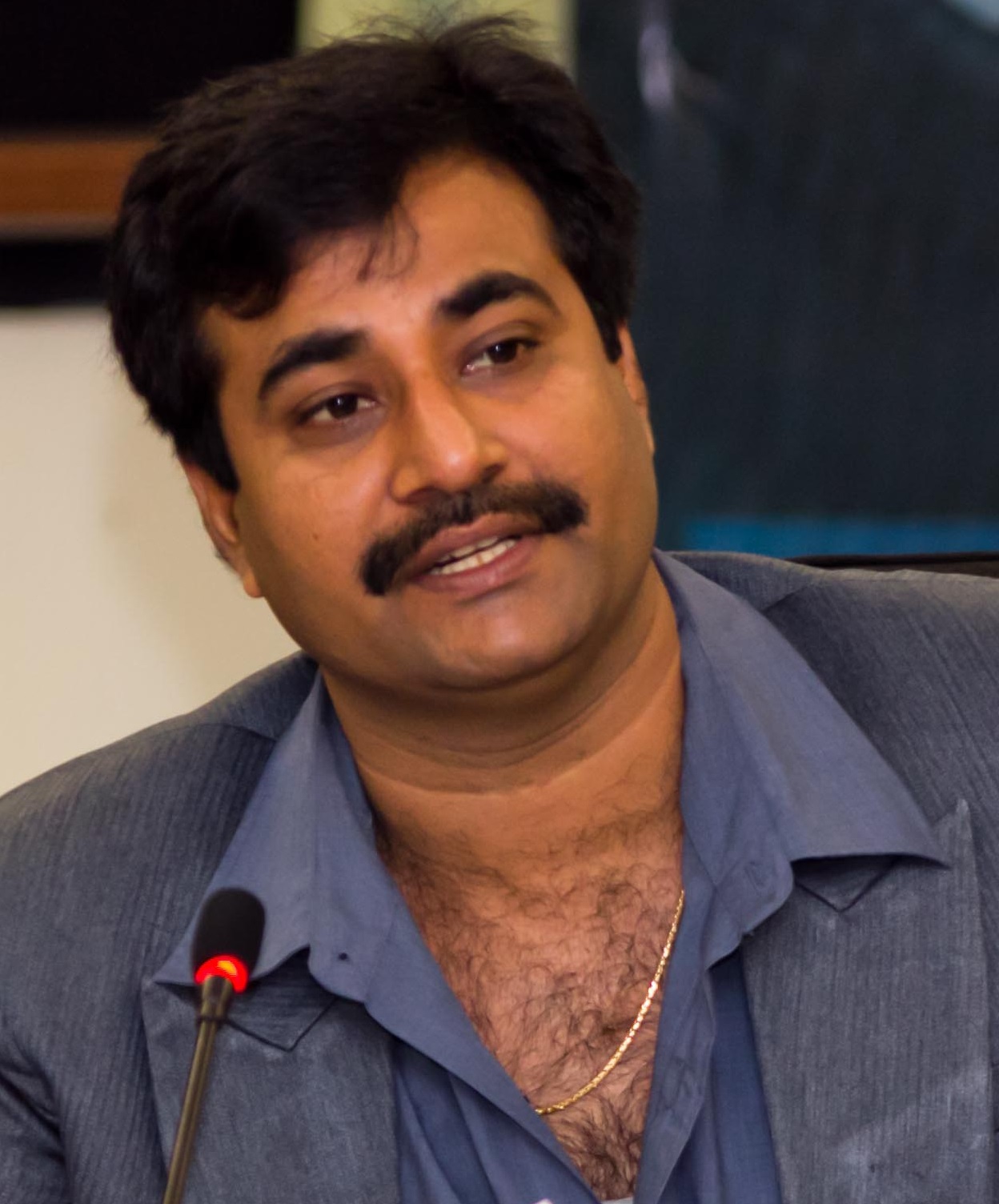}}]{Ashutosh Kumar Singh}
	is working as a Professor and Head in the Department of Computer Applications, National Institute of Technology Kurukshetra, India. He has more than 20 years research and teaching experience in various Universities of the India, UK, and Malaysia. He received his PhD in Electronics Engineering from Indian Institute of Technology, BHU, India and Post Doc from Department of Computer Science, University of Bristol, UK. He is also Charted Engineer from UK. His research area includes Verification, Synthesis, Design and Testing of Digital Circuits, Data Science, Cloud Computing, Machine Learning, Security, Big Data. He has published more than 270 research papers in different journals, conferences and news magazines. 
	
\end{IEEEbiography}
\begin{IEEEbiography}[{\includegraphics[width=0.8\linewidth]{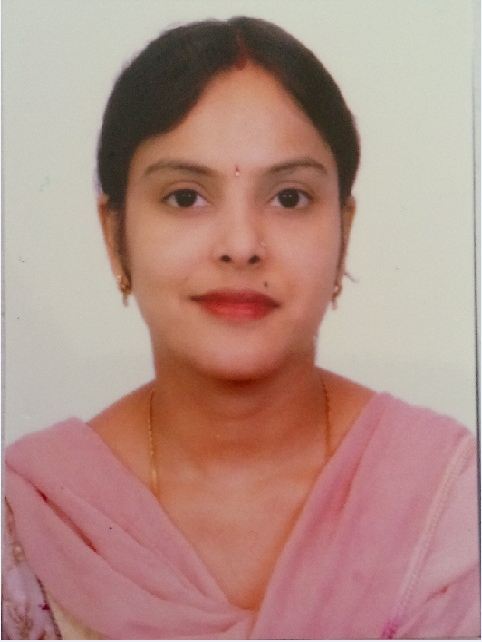}}]{Deepika Saxena}
	  received her M.Tech degree in Computer Science and Engineering  from Kurukshetra University Kurukshetra, Haryana, India in 2014. Currently, she is pursuing her Ph.D from Department of Computer Applications, National Institute of Technology (NIT), Kurukshetra, India. Her major research interests are Neural Networks, Evolutionary Algorithms, Resource Management and Security in Cloud Computing.
\end{IEEEbiography}
\begin{IEEEbiography}[{\includegraphics[width=0.8\linewidth]{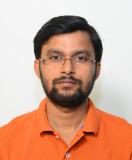}}]{Jitendra Kumar} is an Assistant Professor in the Department of Computer Applications, National Institute of Technology Tiruchirappalli, India. He earned his doctorate from the National Institute of Technology Kurukshetra, India in 2019. His current research interests include Cloud Computing, Machine Learning, Data Analytics, Parallel Processing.
\end{IEEEbiography}

\begin{IEEEbiography}[{\includegraphics[width=0.8\linewidth]{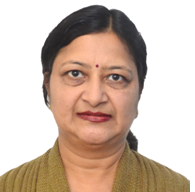}}]{Vrinda Gupta}
received the B.E., M. Tech. and Ph.D. degree in Electronics and Communication Engineering in 1987, 1994 and 2017 respectively, from Nagpur University, Kurukshetra University, and NIT Kurukshetra India. She is currently working as an Associate Professor in the Electronics and Communication Engineering Department at NIT Kurukshetra, India. Her research interests include computer communications, wireless communications \& networking, and wireless sensor networks and IOT.

\end{IEEEbiography}
\end{document}